\documentclass{aa}
\usepackage{graphicx}
\usepackage{txfonts}
\usepackage{ulem}
\usepackage{placeins}

\usepackage[colorlinks=true,citecolor=blue]{hyperref}
\usepackage{natbib}
\begin{document} 
   
   \title{Detection of Na, K, and H$\alpha$ absorption in the atmosphere of WASP-52b using ESPRESSO
   \thanks{Based on observations collected at the European Southern Observatory under ESO programmes 0102.C-0493 and 0102.D-0789.}}

   \author{G. Chen\inst{1}
          \and
          N. Casasayas-Barris\inst{2,3}
          \and
          E. Pall\'{e}\inst{2,3}
          \and
          F. Yan\inst{4}
          \and
          M. Stangret\inst{2,3}
          \and
          H. M. Cegla\inst{5,6}
          \and
          R. Allart\inst{5}
          \and
          C. Lovis\inst{5}
          }

   \institute{Key Laboratory of Planetary Sciences, Purple Mountain Observatory, Chinese Academy of Sciences, Nanjing 210033, China\\
         \email{guochen@pmo.ac.cn}
         \and
             Instituto de Astrof\'{i}sica de Canarias, V\'{i}a L\'{a}ctea s/n, E-38205 La Laguna, Tenerife, Spain
         \and
             Departamento de Astrof\'{i}sica, Universidad de La Laguna, Spain
         \and
             Institut f\"{u}r Astrophysik, Georg-August-Universit\"{a}t, Friedrich-Hund-Platz 1, D-37077 G\"{o}ttingen, Germany
         \and
             Observatoire Astronomique de l’Universit\'{e} de Gen\`{e}ve, chemin des Maillettes 51, 1290 Versoix, Switzerland
         \and
             CHEOPS Fellow, SNSF NCCR-PlanetS}

   \date{Received Month 00, 2019; accepted Month 00, 2019}

 
  \abstract
  {WASP-52b is a low density hot Jupiter orbiting a moderately active K2V star. Previous low-resolution studies have revealed a cloudy atmosphere and found atomic Na above the cloud deck. Here we report the detection of excess absorption at the Na doublet, the H$\alpha$ line, and the K D$_1$ line. We derived a high-resolution transmission spectrum based on three transits of WASP-52b, observed with the ultra stable, high-resolution spectrograph ESPRESSO at the VLT. We measure a line contrast of $1.09\pm 0.16$\% for Na D$_1$, $1.31\pm 0.13$\% for Na D$_2$, $0.86\pm 0.13$\% for H$\alpha$, and $0.46\pm 0.13$\% for K D$_1$, with a line FWHM range of 11--22~km\,s$^{-1}$. We also find that the velocity shift of these detected lines during the transit is consistent with the planet orbital motion, thus confirming their planetary origin. We do not observe any significant net blueshift or redshift that can be attributed to planetary winds. We use activity indicator lines as control but find no excess absorption. However, we do notice signatures arising from the Center-to-Limb variation (CLV) and the Rossiter-McLaughlin (RM) effect at these control lines. This highlights the importance of the CLV+RM correction in correctly deriving the transmission spectrum, which, if not corrected, could resemble or cancel out planetary absorption in certain cases. WASP-52b is the second non-ultra-hot Jupiter to show excess H$\alpha$ absorption, after HD 189733b. Future observations targeting non-ultra-hot Jupiters that show H$\alpha$ could help reveal the relation between stellar activity and the heating processes in the planetary upper atmosphere. }

   \keywords{Planetary systems --
             Planets and satellites: individual: WASP-52b --
             Planets and satellites: atmospheres --
             Techniques: spectroscopic}

   \maketitle
%

\section{Introduction}
\label{sec:intro}

Atomic and ionic species are important diagnostics to characterize exoplanet atmospheres. They carry important information on the heating and cooling processes, chemistry, and dynamics in the upper atmospheres \citep[e.g.,][]{2004Icar..170..167Y,2009ApJ...693...23M,2013ApJ...772..144C,2014ApJ...796...15L,2014ApJ...796...16K,2015ApJ...803L...9H,2017ApJ...851..150H,2018ApJ...855L..11O}. The first detection of an exoplanet atmosphere comes from a space observation of the atomic \ion{Na}{I} line \citep{2002ApJ...568..377C}. Since then, various atoms and ions have been discovered and \ion{Na}{I} is currently the most frequently detected atom in giant planets \citep{2019ARA&A..57..617M}. 

Alkali metal lines, such as \ion{Na}{I} and \ion{K}{I}, can contribute to cooling processes in the upper atmosphere. The development of ground-based high-resolution facilities with resolving power of $\mathcal{R}\sim 10^5$ has enabled transmission spectroscopy to probe upper atmospheres at pressure levels of 10$^{-4}$--10$^{-11}$~bar \citep{2018A&A...612A..53P}. Absorption line profiles of the alkali doublets can be spectrally resolved to infer the temperatures, number densities, and wind patterns at the probed atmospheric layers \citep[e.g.,][]{2015ApJ...803L...9H,2015A&A...577A..62W,2020A&A...633A..86S}, which makes high-resolution spectroscopy highly competitive even in the era of {\it James Webb} space telescope ($\mathcal{R}\sim 10^3$). 
The first ground-based detections of \ion{Na}{I} were achieved by high-resolution transmission spectroscopy in the hot Jupiters HD 189733b \citep{2008ApJ...673L..87R} and HD 209458b \citep{2008A&A...487..357S}. Only until recently has \ion{K}{I} been detected by high-resolution spectroscopy \citep[HD 189733b;][]{2019MNRAS.489L..37K}. However, \citet{CasasayasBarris2020} found no detectable \ion{Na}{I} absorption in HD 209458b based on five transits acquired with the HARPS-N and CARMENES spectrographs, revealing that the Center-to-Limb variation (CLV) and the Rossiter-McLaughlin \citep[RM;][]{1924ApJ....60...15R,1924ApJ....60...22M} effect could strongly affect the search of planetary absorption.

On the other hand, H$\alpha$ is a sensitive probe of heating  because its formation in the planetary atmospheres requires strict local conditions \citep[e.g., local particle densities, temperature, and radiation field;][]{2017ApJ...851..150H}. H$\alpha$ was first detected in the classical hot Jupiter HD 189733b \citep{2012ApJ...751...86J,2016AJ....152...20C,2017AJ....153..217C,2017AJ....153..185C}, which shows variable absorption depths that may be due to the variability in the stellar Lyman continuum \citep{2017AJ....153..217C,2017AJ....153..185C,2017ApJ...851..150H}. Recently, the newly emerged class of exoplanets -- ultra-hot Jupiters (UHJs), which are gas giants with dayside temperatures typically greater than 2200 K \citep{2018A&A...617A.110P} -- offer a new avenue for the study of H$\alpha$ absorption. Four UHJs, i.e., KELT-20b \citep{2018A&A...616A.151C,2019A&A...628A...9C}, KELT-9b \citep{2018NatAs...2..714Y,2019AJ....157...69C,2020ApJ...888L..13T}, WASP-12 \citep{2018AJ....156..154J}, and WASP-121b \citep{2020arXiv200107196C}, have been found to show H$\alpha$ absorption, unassociated with stellar variability. This increasing H$\alpha$-host sample has different stellar environments \citep[e.g.,][]{2018ApJ...868L..30F}, but intriguingly, all of them also show excess absorption at \ion{Na}{I} doublet, which, together with atomic lines like \ion{He}{I} 1083~nm \citep[e.g.,][]{2018Sci...362.1388N,2018Sci...362.1384A}, holds a great opportunity to understand atmospheric heating and cooling with multi-tracers. 

Here we report the second non-UHJ planet to show excess H$\alpha$ absorption. WASP-52b is a low density hot Jupiter orbiting a moderately active K2 dwarf ($\log R^\prime_\mathrm{HK}=-4.4$, \citealt{2013A&A...549A.134H}; $\log L_\mathrm{x}/L_\mathrm{bol}=-4.7$, \citealt{2018AJ....156..298A}) that has shown occulted starspots \citep{2017MNRAS.465..843M} and faculae \citep{2016MNRAS.463.2922K} in the transit light curves. \citet{2017A&A...600L..11C} found that WASP-52b has a cloudy atmosphere, with a detection of a narrow Na core and a tentative detection of the K doublet above the cloud deck. The Na detection was later confirmed by {\it Hubble} space telescope observations \citep{2018AJ....156..298A}.

This paper is organized as follows: we summarize the observations in Sect.~\ref{sec:data}, and detail the methods to derive transmission spectrum in Sect.~\ref{sec:analysis}. We search for excess absorption at lines of interest in Sect.~\ref{sec:results} and discuss the obtained results in Sect.~\ref{sec:discuss}. Conclusions are given in Sect.~\ref{sec:conclusions}.

\section{Observations and data reduction}
\label{sec:data}

\begin{table*}[ht!]
     \small
     \centering
     \caption{Summary of the ESPRESSO observations.}
     \label{tab:obslog}
     \begin{tabular}{cccccccccccc}
     \hline\hline\noalign{\smallskip}
     \# & Calendar   & UT obs. & $T_\mathrm{exp}$ & $N_\mathrm{obs}$ (in/out) & Airmass\tablefootmark{a} & \multicolumn{3}{c}{S/N\tablefootmark{b}} & Telescope & PID\tablefootmark{d}\\\noalign{\smallskip}
     \cline{7-9}\noalign{\smallskip}
     & Night & window   & [s] & & & Continuum\tablefootmark{c} & Na core & H$\alpha$ core & &\\\noalign{\smallskip}
     \hline\noalign{\smallskip}
     1  & 2018-10-31 & 00:51-04:37 & 500 & 24 (13/11) & 1.21--1.20--1.98 & 34--46 & 2--7  & 13--19 & VLT-UT3 & i\\\noalign{\smallskip}
     2  & 2018-11-07 & 00:05-04:24 & 800 & 18 (9/9) & 1.22--1.20--2.17   & 38--57 & 3--10 & 15--24 & VLT-UT3 & ii\\\noalign{\smallskip}
     3  & 2018-11-14 & 00:23-04:13 & 800 & 16 (8/8) & 1.20--1.20--2.46         & 37--61 & 4--10 & 15--26 & VLT-UT2 & ii\\\noalign{\smallskip}
     \hline\noalign{\smallskip}
     \end{tabular}
     \tablefoot{
     \tablefoottext{a}{The first and third values refer to the airmass at the beginning and at the end of the observation. The second value gives the minimum airmass.}
     \tablefoottext{b}{The two figures correspond to the minimum and maximum S/N, respectively.}
     \tablefoottext{c}{The S/N of the continuum was measured at around 590~nm.}
     \tablefoottext{d}{PID i and ii refer to ESO programs 0102.D-0789 (PI: H. Cegla) and 0102.C-0493 (PI: G. Chen), respectively.}
     }
\end{table*}

We observed three transits of the hot Jupiter WASP-52b under the ESO programs 0102.C-0493 (PI: G. Chen) and 0102.D-0789 (PI: H.~M. Cegla), using the ultra-stable fibre-fed \'{e}chelle high-resolution spectrograph ESPRESSO \citep{2010SPIE.7735E..0FP}, installed at the incoherent combined Coud\'{e} facility of the VLT. We employed the High Resolution 1-UT mode and configured the CCD in 2$\times$1 binning and slow readout mode. This gives a spectral resolving power of $R$$\sim$140\,000 and covers a wavelength range of 380--788~nm. During the observations, fiber A was pointed to the target star, while fiber B was pointed to sky for a simultaneous monitoring of the sky emission. All the three transits were observed in photometric conditions. A summary observing log is given in Table~\ref{tab:obslog}.

We reduced the raw spectral images using the ESPRESSO data reduction pipeline (version 1.3.2)\footnote{See \url{ftp://ftp.eso.org/pub/dfs/pipelines/instruments/espresso/espdr-pipeline-manual-1.3.2.pdf}.}. The data were corrected for bias, dark, flat, and bad pixels. The extracted spectra were deblazed, slice-merged, and order-merged. The sky spectra measured in fiber B were scaled to account for the fiber-to-fiber relative efficiency and subtracted from the science spectra measured in fiber A. We further rebinned the ESPRESSO pipeline 1D spectra into wavelength intervals of 0.01~$\AA$ using the IDL routine \texttt{rebinw} from the \texttt{PINTofALE} package \citep{2010ascl.soft07001D} for flux conservation.

We coadded the out-of-transit stellar spectra of WASP-52, and used the \texttt{zaspe} code \citep{2017MNRAS.467..971B} to improve the estimation of its stellar atmospheric parameters. We obtained an effective temperature of $T_\mathrm{eff}=5014\pm 41$~K, a stellar surface gravity of $\log g_\star=4.48\pm 0.08$, a metallicity of $\mathrm{[Fe/H]}=0.06\pm 0.03$, and a projected rotation velocity of $v\sin i_\star=3.13\pm 0.42$~km\,s$^{-1}$. The revised parameters are consistent with those of \citet{2013A&A...549A.134H}, but with smaller uncertainties.

\begin{table}[ht!]
     \small
     \centering
     \caption{Physical and orbital parameters of the WASP-52 system.}
     \label{tab:param}
     \begin{tabular}{llc}
     \hline\hline\noalign{\smallskip}
     Parameter & Symbol [Unit] & Value \\\noalign{\smallskip}
     \hline\noalign{\smallskip}
     \multicolumn{3}{c}{\dotfill\it Stellar Parameters\dotfill}\\\noalign{\smallskip}
       Stellar mass            & $M_\star$ [$M_\sun$]            & 0.87 $\pm$ 0.03       \tablefootmark{a}     \\
       Stellar radius          & $R_\star$ [$R_\sun$]            & 0.79 $\pm$ 0.02       \tablefootmark{a}     \\
       Effective temperature   & $T_\mathrm{eff}$ [K]      & 5014 $\pm$ 41      \tablefootmark{d}     \\
       Surface gravity         & $\log g_\star$ [cgs]       & 4.48 $\pm$ 0.08     \tablefootmark{d}     \\
       Metallicity             & $\mathrm{[Fe/H]}$ [dex]     & 0.06 $\pm$ 0.03   \tablefootmark{d}     \\
       Projected rotation velocity        & $v\sin i_\star$ [$\mathrm{km\,s}^{-1}$]  & 3.13 $\pm$ 0.42      \tablefootmark{d}     \\
     \multicolumn{3}{c}{\dotfill\it Planet Parameters\dotfill}\\\noalign{\smallskip}
        Planet mass             & $M_p$ [$M_J$]                & 0.46 $\pm$ 0.02          \tablefootmark{a}     \\
        Planet radius           & $R_p$ [$R_J$]                & 1.27 $\pm$ 0.03          \tablefootmark{a}     \\
        Equilibrium temperature & $T_\mathrm{eq}$ [K]        & 1315 $\pm$ 35              \tablefootmark{a}     \\
        Surface gravity         & $\log g_p$ [cgs]           & 2.81 $\pm$ 0.03            \tablefootmark{a}     \\
     \multicolumn{3}{c}{\dotfill\it System Parameters\dotfill}\\\noalign{\smallskip}
        Transit epoch           & $T_0$ [BJD$_\mathrm{TDB}$]                & 2456862.79776  \tablefootmark{b}     \\
        Orbital period          & $P$ [d]                 & 1.74978119         \tablefootmark{b}     \\
        Planet-to-star radius ratio  & $R_p/R_\star$    & 0.1608 $\pm$ 0.0018           \tablefootmark{c}     \\
        Scaled semi-major axis  & $a/R_\star$           & 7.14 $\pm$ 0.12               \tablefootmark{c}     \\
        Orbital inclination     & $i$ [deg]                  & 85.06 $\pm$ 0.27          \tablefootmark{c}     \\
        Transit duration        & $T_{14}$ [d]             & 0.07734         \tablefootmark{c}     \\
        Fully in-transit duration & $T_{23}$ [d]           & 0.04477         \tablefootmark{c}     \\
        Eccentricity            & $e$                   & 0 (fixed)                     \tablefootmark{a}     \\
        Stellar RV semi-amplitude    & $K_\star$ [$\mathrm{km\,s}^{-1}$]    & 0.0843 $\pm$ 0.0030    \tablefootmark{a}     \\
      \multicolumn{3}{c}{\dotfill\it Rossiter-McLaughlin Parameters\dotfill}\\\noalign{\smallskip}
        Systemic velocity \#1       & $\gamma$ [$\mathrm{m~s}^{-1}$]   & $-$821.16 $\pm$ 0.66        \tablefootmark{e} \\
        Systemic velocity \#2       & $\gamma$ [$\mathrm{m~s}^{-1}$]   & $-$843.39 $\pm$ 0.63        \tablefootmark{e} \\
        Systemic velocity \#3       & $\gamma$ [$\mathrm{m~s}^{-1}$]   & $-$835.69 $\pm$ 0.56        \tablefootmark{e} \\
        Joint offset of mid-transit        & $\Delta T_\mathrm{C}$ [d]  & 0.00031 $\pm$ 0.00032       \tablefootmark{e} \\
        Stellar RV semi-amplitude    & $K_\star$ [$\mathrm{km\,s}^{-1}$]    & 0.0828 $\pm$ 0.0021    \tablefootmark{e}     \\
        Projected rotation velocity        & $v\sin i_\star$ [$\mathrm{km\,s}^{-1}$]      & 2.622 $\pm$ 0.074       \tablefootmark{e}     \\
        Projected spin-orbit angle     & $\lambda$ [deg]            & 1.1 $\pm$ 1.1   \tablefootmark{e}     \\
        Limb darkening coefficient       & $u_1$ & 0.94 $\pm$ 0.06       \tablefootmark{e}     \\
    \hline\noalign{\smallskip}
    \end{tabular}
    \tablefoot{
      \tablefoottext{a}{\citet{2013A&A...549A.134H}.}
      \tablefoottext{b}{\citet{2017MNRAS.465..843M}.}
      \tablefoottext{c}{\citet{2017A&A...600L..11C}.}
      \tablefoottext{d}{This work (spectroscopic analysis).}
      \tablefoottext{e}{This work (Rossiter-McLaughlin analysis).}
    }
\end{table}

\section{Methods}
\label{sec:analysis}

We carried out the following steps to search for absorption signatures originating in the planetary atmosphere.

\subsection{Determination of in- or out-of-transit}
\label{sec:inout}

We calculated the mid-transit times of the three nights based on literature ephemeris listed in Table~\ref{tab:param}, which were further refined by fitting for the Rossiter-McLaughlin (RM) effect. The fitting was performed with the EXOFAST \citep{2013PASP..125...83E} parameterization of the \citet{2005ApJ...622.1118O} formulae, and the details can be found in Appendix \ref{sec:rmfit}. The out-of-transit frames are strictly defined as the exposures that have zero overlap with the transit event. The fully in-transit frames are defined as the exposures that have a fraction more than $\sim$50\% located between the second and third contacts of the transit event. The remaining frames are assigned as either ingress or egress. 

\begin{figure}
\centering
\includegraphics[width=\linewidth]{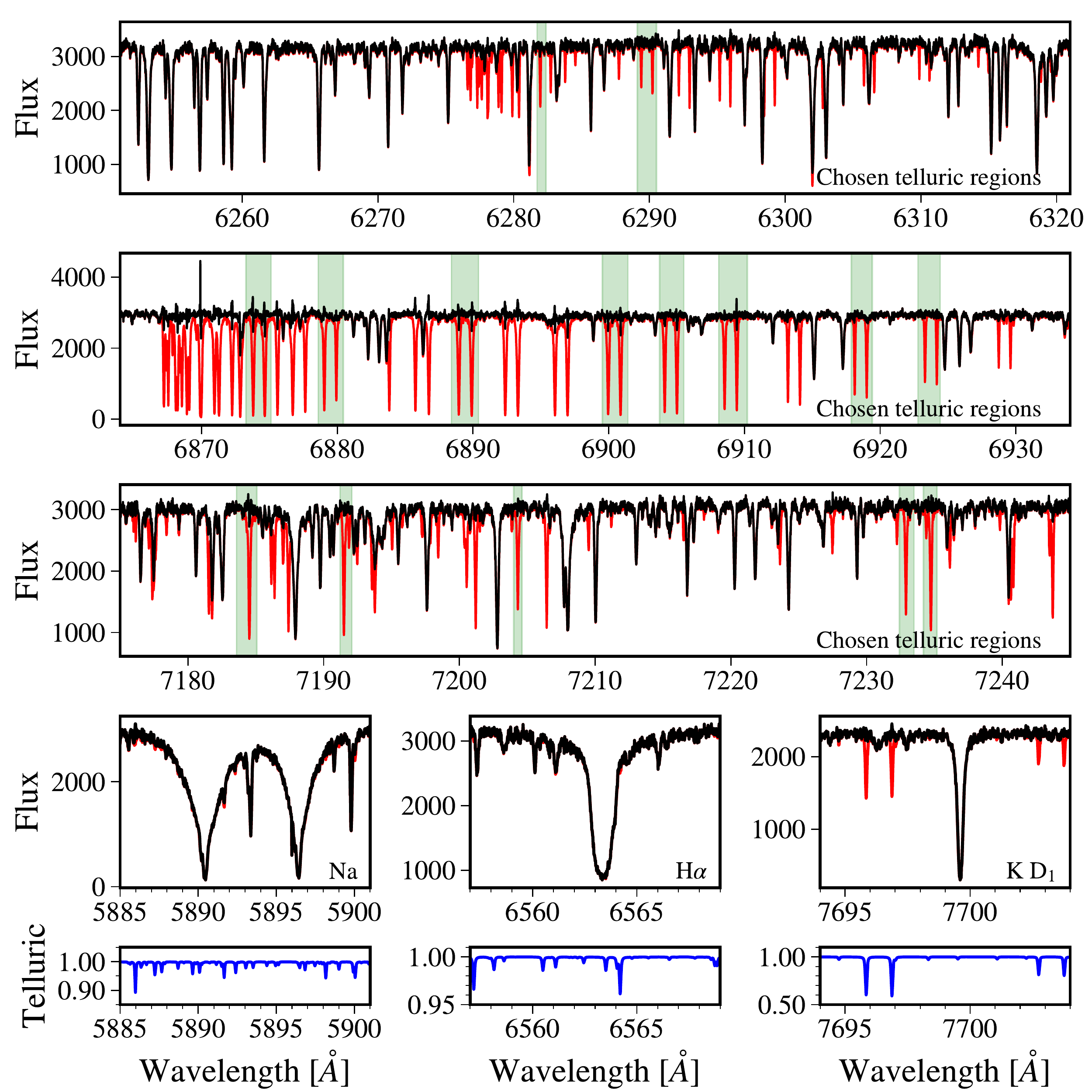}
\caption{Illustration of telluric correction by \texttt{molecfit}. The top four rows show the example spectrum before (red line) and after (black line) the telluric correction. The top three rows present the fifteen telluric regions (green shaded areas) to fit for the telluric absorption. The fourth row present the correction at the Na, H$\alpha$, and K lines. The bottom row shows the telluric correction applied in the fourth row.}
\label{fig:telluric_correction}
\end{figure}

\subsection{Removal of telluric absorption}
\label{sec:telluric}

While the use of fiber B on sky has helped remove the telluric emission lines (e.g., telluric Na), further correction of telluric absorption in the stellar spectra is necessary. We used the ESO software \texttt{molecfit} version 1.5.7 \citep{2015A&A...576A..77S,2015A&A...576A..78K} to perform telluric correction, based on a synthetic modeling of the Earth's atmospheric transmission with a line-by-line radiative transfer model. This approach has been adopted in several recent studies of high-resolution transmission spectroscopy \citep{2017A&A...606A.144A,2018Sci...362.1384A,2019A&A...623A..58A,2018Sci...362.1388N,2018A&A...620A..97S,2019A&A...623A.166S,2019A&A...628A...9C,2019A&A...627A.165H,2019A&A...629A.110A,2019AJ....157...69C,2020arXiv200107196C,2020arXiv200107667K}. 

We shifted the pipeline spectra back to the terrestrial rest frame before the use of \texttt{molecfit}, because they have been corrected for the Barycentirc Earth Radial Velocity (BERV). We carefully selected fifteen telluric wavelength regions, spreading from 6283~$\AA$ to 7238~$\AA$, to fit for the telluric absorption. These regions contain only unsaturated and separated strong lines of telluric H$_2$O or O$_2$, free of stellar lines. The best-fit solution was applied to the whole wavelength range of ESPRESSO. Fig.~\ref{fig:telluric_correction} shows the selected fifteen telluric wavelength regions and illustrates the telluric correction performed by \texttt{molecfit}. 

\subsection{Removal of stellar lines}
\label{sec:stellarline}

After telluric correction, we shifted all the spectra to the stellar rest frame to remove the stellar lines, which considered both BERV  and stellar RV. The RV model adopts the best-fit model obtained in Appendix \ref{sec:rmfit} but without the RM component. It also corrected the systemic velocity in each individual night. 

The removal of stellar lines relied on the out-of-transit master frame (hereafter master-out), and it was performed in each night individually. We first normalized each out-of-transit spectrum by its median flux and then median-combined them to create a preliminary master-out, which served as a reference to correct possible continuum variation. We corrected the difference between any individual spectrum and the preliminary master-out by fitting the individual-to-master-out ratio spectrum with a fourth-order polynomial function.  The best-fit polynomial model was removed from each individual spectrum. Finally, we combined the corrected out-of-transit spectra to create the final master-out, with the inverse square of flux uncertainties as the weight. We divided each corrected individual spectrum by this master-out to remove stellar lines. We further normalized each residual spectrum by the continuum outside of the planetary absorption lines of interest.

\begin{figure*}
\centering
\includegraphics[width=1\linewidth]{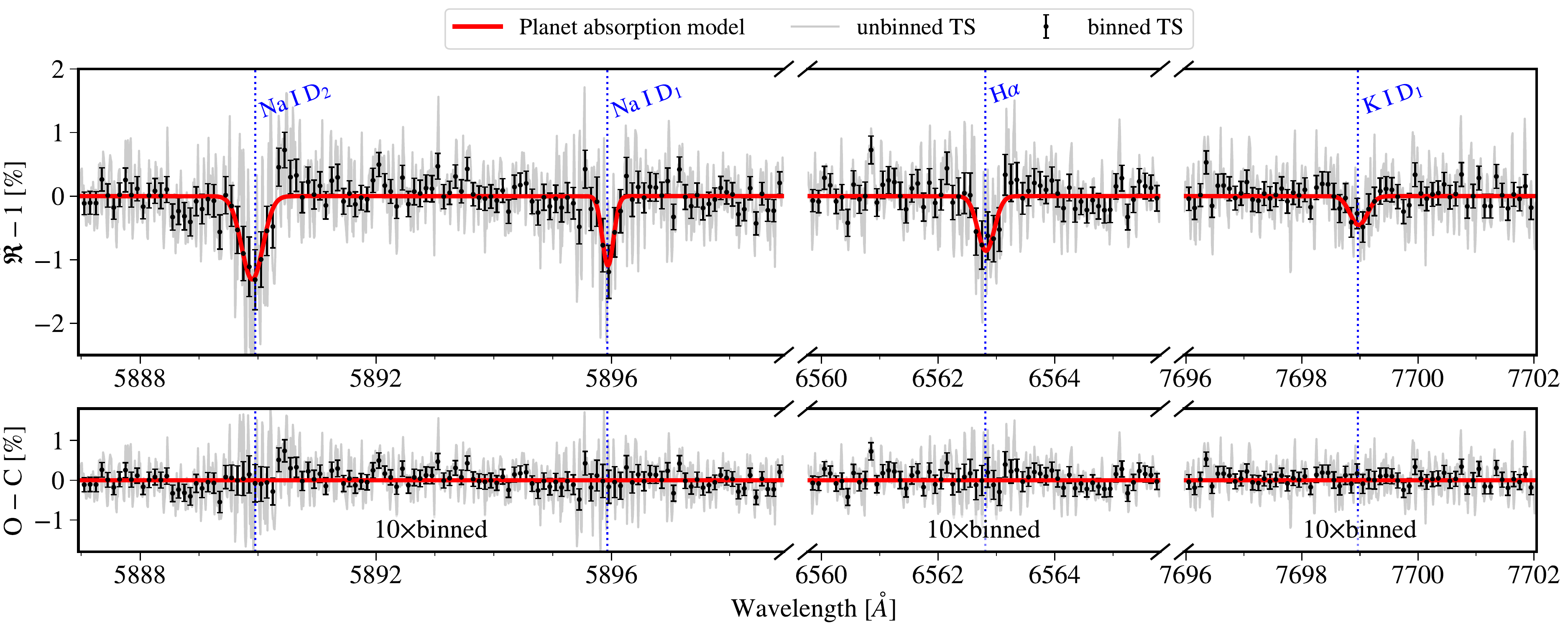}
\caption{{\it Top row:} transmission spectrum of WASP-52b observed by ESPRESSO at the Na doublet, H$\alpha$, and K D$_1$ lines. Signatures arising from the center-to-limb variation and Rossiter-McLaughlin effects have been corrected. The gray line shows the unbinned transmission spectrum in grids of 0.01~$\AA$. The black circles are binned into intervals of 0.1~$\AA$ (i.e., 10$\times$binned). The red line shows the best-fit planet absorption model. The error bars show the propagated photon noise. {\it Bottom row:} the best-fit residuals.}
\label{fig:transpec}
\end{figure*}

\subsection{Correction of stellar center-to-limb variation and Rossiter-McLaughlin effect}

The transit of a planet not only potentially imprints planetary absorption signals on the stellar lines, but also reveals stellar line deformation of different origins that could complicate the search of planetary absorption signals. Two important contaminations that will introduce deformation to the stellar lines during the transit are the center-to-limb variation (CLV) and the RM effect \citep[e.g.][]{2015A&A...574A..94Y,2017A&A...603A..73Y,2015A&A...582A..51C,2015ApJ...814L..24L,2016ApJ...819...67C}. We followed the approach detailed in \citet{2019A&A...628A...9C} to generate the CLV+RM model. This approach uses the \texttt{Spectroscopy Made Easy} tool \citep[SME;][]{1996A&AS..118..595V,2017A&A...597A..16P} to compute the stellar spectra at 21 different angles ($\mu=\cos\theta$). The calculation uses the VALD3 \citep{2015PhyS...90e4005R} line lists and Kurucz ATLAS9 models. Local thermodynamic equilibrium (LTE) is adopted. The stellar disk is divided into cells of $0.01R_\star\times 0.01R_\star$ and each cell has its own spectrum interpolated from the pre-calculated grids based on its projected rotation velocity and $\mu$ value. At a given orbital phase, the spectra of unobscured cells are integrated to derive the final model spectrum. Consequently, the time-series model spectra contain both CLV and RM effects. Due to limited S/N, we assumed the white-color planet radius $R_\mathrm{p}$ as the radius of the obscuring disk, without considering an additional annulus that could be introduced by the planetary atmosphere at given lines. This CLV+RM model data cube is normalized by the out-of-transit model spectrum that has no CLV+RM effects, and then corrected in the residual spectrum matrix obtained in Sect.~\ref{sec:stellarline}.

\subsection{Creation of transmission spectrum}

After the CLV+RM correction, we shifted all the in-transit residual spectra to the planet-rest frame using $v_\mathrm{p}=K_\mathrm{p}\sin 2\pi\phi$, where $\phi$ is the orbital phase and $K_\mathrm{p}=K_\star M_\star/M_\mathrm{p}=167\pm 11$~km\,s$^{-1}$ is the expected planet RV semi-amplitude based on the law of conservation of momentum. We then weight-combined all the fully in-transit spectra and normalized the continuum outside the regions of interest to create the final transmission spectrum $\tilde{\mathfrak{R}}$. 

\begin{table*}
  \centering
  \caption{Parameters from the Gaussian fit to the line profile. }
  \label{tab:line}
  \begin{tabular}{cccc|ccc}
  \hline\hline\noalign{\smallskip}
  Line & Contrast & Center  & FWHM    & Effective radius  & Center offset & FWHM\\
       & [\%]     & [$\AA$] & [$\AA$] & [$R_\mathrm{p}$]  & [km\,s$^{-1}$]  & [km\,s$^{-1}$]\\
  \hline\noalign{\smallskip}
Na D$_1$  & 1.09 $\pm$ 0.16 & 5895.937 $\pm$ 0.018 & 0.226 $\pm$ 0.032 & 1.193 $\pm$ 0.026 & $+$0.6 $\pm$ 0.9 & 11.5 $\pm$ 1.6\\
Na D$_2$  & 1.31 $\pm$ 0.13 & 5889.900 $\pm$ 0.024 & 0.424 $\pm$ 0.036 & 1.228 $\pm$ 0.021 & $-$2.6 $\pm$ 1.2 & 21.6 $\pm$ 1.8\\
H$\alpha$ & 0.86 $\pm$ 0.13 & 6562.821 $\pm$ 0.029 & 0.336 $\pm$ 0.039 & 1.155 $\pm$ 0.022 & $+$0.5 $\pm$ 1.3 & 15.4 $\pm$ 1.8\\
K D$_1$   & 0.46 $\pm$ 0.13 & 7698.979 $\pm$ 0.050 & 0.352 $\pm$ 0.084 & 1.085 $\pm$ 0.023 & $+$0.6 $\pm$ 1.9 & 13.7 $\pm$ 3.3\\
  \hline\noalign{\smallskip}
  \end{tabular}
  \tablefoot{
      The last three columns are derived from the parameters in the second to fourth columns.
    }
\end{table*}

\section{Results}
\label{sec:results}

\subsection{Spectrally resolved Na doublet, H$\alpha$, and K D$_1$}
\label{sec:detection}

We have searched ESPRESSO's full wavelength coverage for spectral lines that may originate in the planetary atmosphere, and only found excess absorption at Na doublet, H$\alpha$, and K D$_1$. We were not able to derive reliable transmission spectrum at K D$_2$ because it is blended with two strong telluric oxygen lines. We fitted a Gaussian function to each spectral line. The fit was performed on the unbinned transmission spectrum. The associated uncertainties were derived using the posterior distribution from the ``prayer-bead'' approach \citep[e.g.,][]{2008MNRAS.386.1644S}, in which, the best-fit residuals were sequentially shifted, added back to the best-fit model, and then fitted by the Gaussian function again. We present the parameters of the line profiles, including contrast, center, and FWHM (full width at half maximum), in Table~\ref{tab:line}. We show the CLV+RM-corrected transmission spectrum of WASP-52b zoomed at Na, H$\alpha$, and K D$_1$ in Fig.~\ref{fig:transpec}.

The measured line centers of these detected lines are consistent with their rest-frame wavelengths (Na D$_1$: 5895.924~$\AA$, Na D$_2$: 5889.951~$\AA$, H$\alpha$: 6562.81~$\AA$, K D$_1$: 7698.965~$\AA$). The weighted average of velocity shift is $-0.2\pm 0.6$~km\,s$^{-1}$, indicating that no net blueshift/redshift is observed. The measured FWHM in units of velocity are $11.5\pm 1.6$~km\,s$^{-1}$ (Na D$_1$), $21.6\pm 1.8$~km\,s$^{-1}$ (Na D$_2$), $15.4\pm 1.8$~km\,s$^{-1}$ (H$\alpha$), $13.7\pm 3.3$~km\,s$^{-1}$ (K D$_1$), respectively. The line contrast ratio $\sim$1.2 and FWHM of Na doublet are very similar to those of WASP-49b \citep[$h_{D_2}/h_{D_1}\sim1.09$, $11.1\pm 4.3$~km\,s$^{-1}$ for D$_1$, $21.3\pm 4.9$~km\,s$^{-1}$ for D$_2$;][]{2017A&A...602A..36W}, which is a hot Jupiter of similar bulk properties and equilibrium temperature to WASP-52b, with a cloudy atmosphere as well \citep{2016A&A...587A..67L,2017ApJ...849..145C}. 

We also calculated the effective planet radius at the line center assuming $R_\mathrm{eff}^2/R_\mathrm{p}^2=(\delta +h)/\delta$, where $\delta=(R_\mathrm{p}/R_\star)^2$ is obtained from Table~\ref{tab:param} and $h$ is the line contrast. The derived effective radius is $1.193\pm 0.026$~$R_\mathrm{p}$ for Na D$_1$, $1.228\pm 0.021$~$R_\mathrm{p}$ for Na D$_2$, $1.155\pm 0.022$~$R_\mathrm{p}$ for H$\alpha$, and $1.085\pm 0.023$~$R_\mathrm{p}$ for K D$_1$.

\subsection{Binned absorption depth}
\label{sec:binned_ad}

\begin{figure}
\centering
\includegraphics[width=\linewidth]{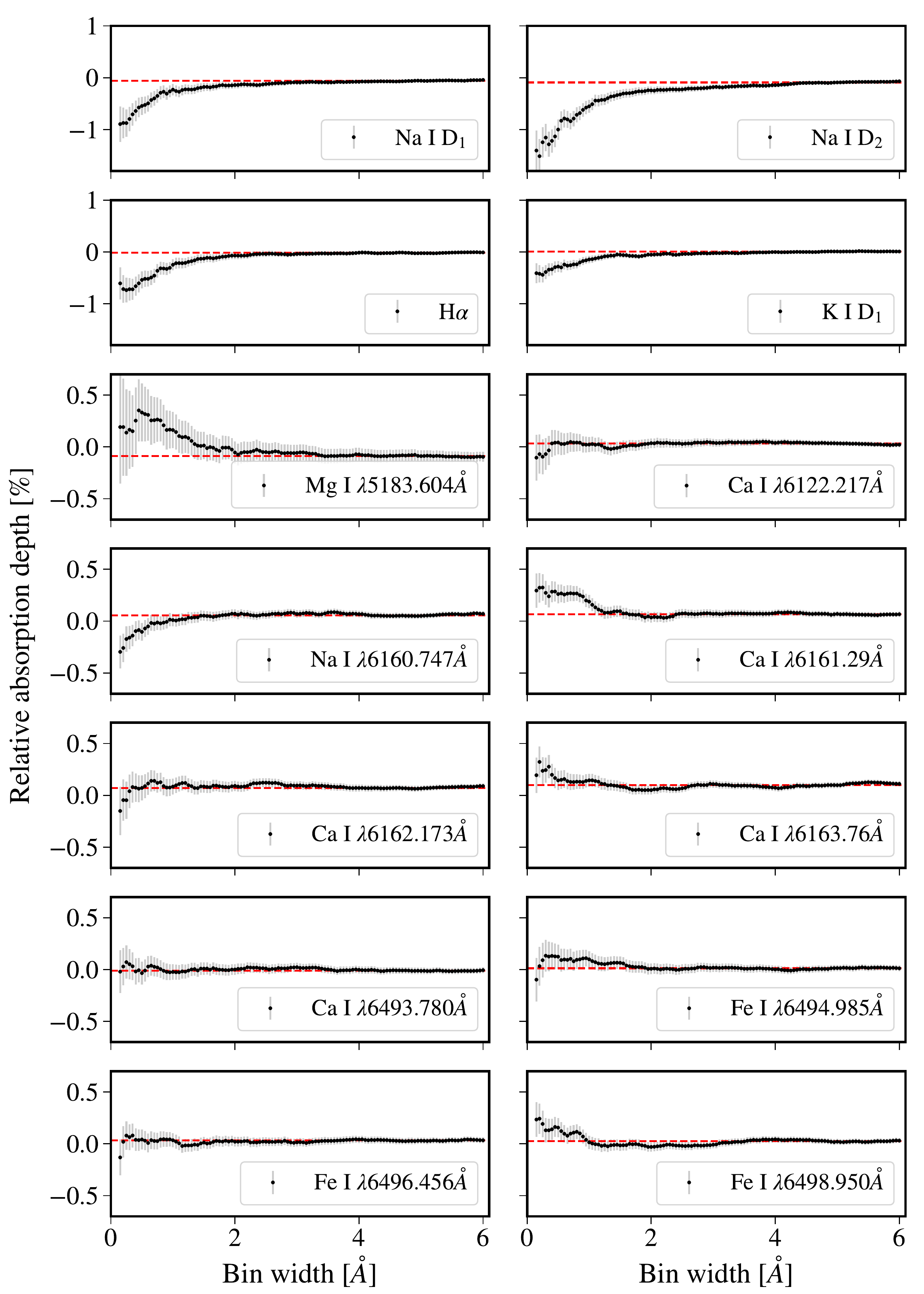}
\caption{Absorption depth as a function of bin width, which is measured in a band centered at the target line. Fourteen lines are shown, including the Na doublet, H$\alpha$, K D$_1$ lines and ten stellar activity indicator control lines. The top two rows have a different y-axis scale from the remaining rows. The dashed line shows the average value for bin widths larger than 4~$\AA$.}
\label{fig:adgrowth}
\end{figure}

A more general way to evaluate the strength of the excess absorption at the target line is the so-called binned absorption depth, which is widely adopted in both low-to-mid resolution observations \citep[e.g.,][]{2002ApJ...568..377C,2008ApJ...686..658S} and high resolution observations \citep[e.g.,][]{2008A&A...487..357S,2008ApJ...673L..87R,2015A&A...577A..62W}. In our case, we derived the binned absorption depth by averaging the final transmission spectrum over a passband centered at the target line, which has already been corrected for the CLV+RM effects. We note that the Gaussian fit presented in Sect.~\ref{sec:detection} only works if the passband bin width is sufficiently narrow to well resolve the line. Therefore, the binned absorption depth delivers the strength of the excess absorption within a passband bin at the line center without the need for resolving the line, which allows a straightforward comparison between different planets observed at different resolutions. 

We varied the width of the passband in steps of 0.05~$\AA$ and found that the absorption depth achieves the best S/N within the 0.4~$\AA$ band for Na doublet, 0.35~$\AA$ band for H$\alpha$, and 0.55~$\AA$ band for K D$_1$. We measured $0.90\pm 0.15$~\% (2$\times$0.4~$\AA$ Na band), $0.72\pm 0.20$~\% (0.35~$\AA$ H$\alpha$ band), and $0.30\pm 0.10$~\% (0.55~$\AA$ K D$_1$ band), respectively. The best S/N bin width is in broad agreement with the FWHM derived in Sect.~\ref{sec:detection}. 

Fig.~\ref{fig:adgrowth} presents the growth curve of the binned absorption depth at the Na, H$\alpha$, and K D$_1$ lines, together with a selection of stellar activity indicator lines. The latter serves as a control, and will be discussed in Sect.~\ref{sec:control}. The planetary lines and the control lines exhibit different shape of growth curves. While the planetary lines have a well shaped absorption profile, the control lines are approximately consistent with zero, showing that we have only detected excess absorption at the Na, H$\alpha$, and K D$_1$ lines. 

\citet{2017A&A...600L..11C} obtained an absorption depth of $\Delta F/F=0.378\pm0.068$\% in a 16~$\AA$ band at Na with GTC low-resolution transmission spectrum, where the Na doublet is unresolved. For comparison, we integrated the ESPRESSO transmission spectrum over a 16-$\AA$ passband centered at the Na doublet ($A\approx 0.62$\%), and convolved it with a Gaussian kernel. The kernel has a FWHM of 10~$\AA$ (i.e. seeing-limited resolution in the GTC observations). This would result in an absorption depth of $\sim$0.385\% in a 16~$\AA$ band, which is consistent with the GTC measurement.

\begin{figure}
\centering
\includegraphics[width=\linewidth]{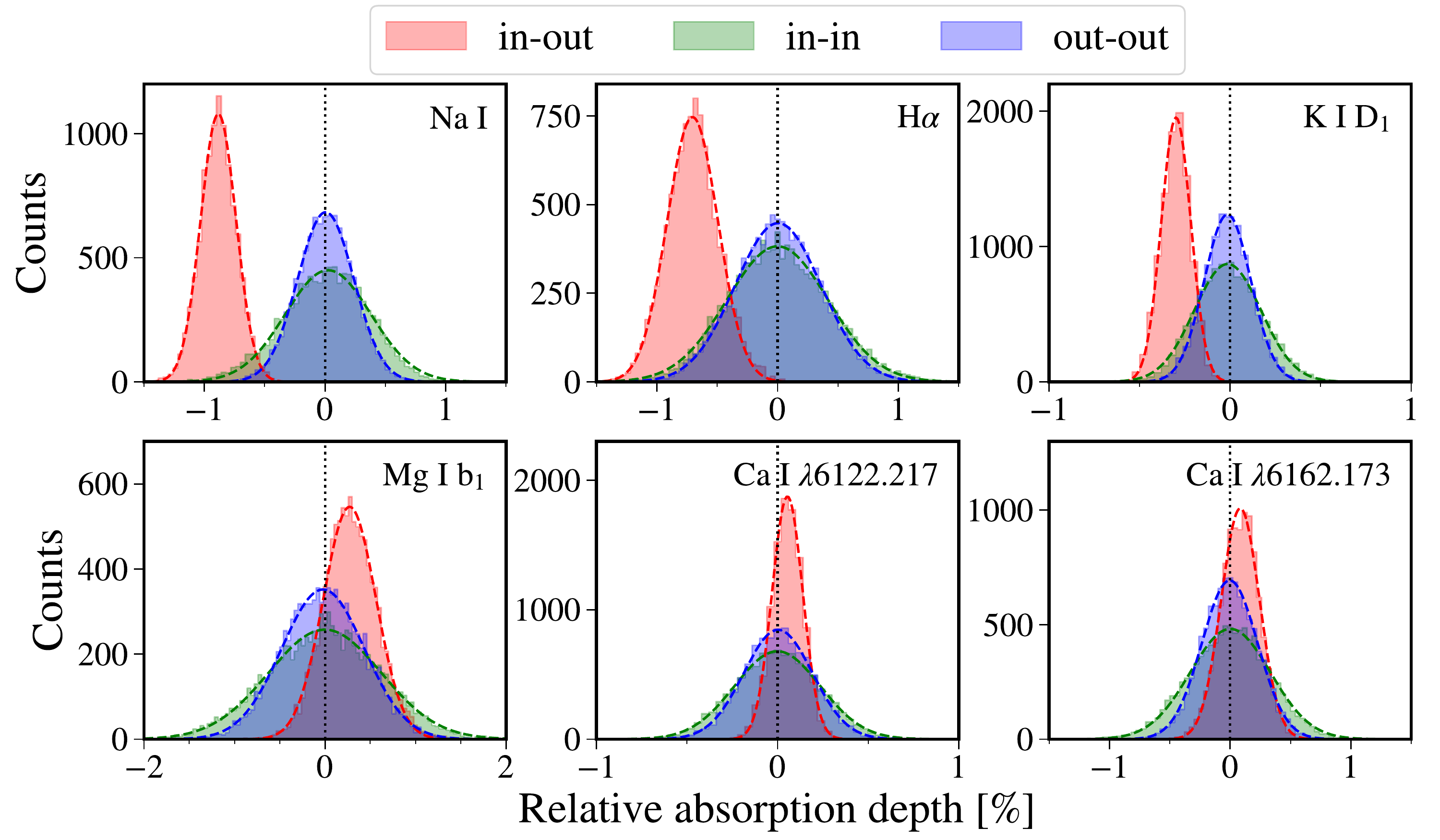}
\caption{Distribution of absorption depths from the empirical Monte Carlo (EMC) simulation at the Na doublet, H$\alpha$, and K D$_1$ lines and three control lines (\ion{Mg}{I} 5183.604$\AA$, \ion{Ca}{I} 6122.217$\AA$, \ion{Ca}{I} 6162.173$\AA$). The control lines are sensitive to stellar activity. The absorption depths are measured in two 0.4~$\AA$ passbands for Na doublet, a 0.35~$\AA$ passband for H$\alpha$, a 0.55~$\AA$ passband for K D$_1$, and a 0.4~$\AA$ passband for the three control lines, respectively. The red, green, and blue distributions correspond to the ``in-out'', ``in-in'', and ``out-out'' scenarios detailed in Sect.~\ref{sec:emc}, respectively. }
\label{fig:emc}
\end{figure}

\begin{figure*}
\centering
\includegraphics[width=0.82\linewidth]{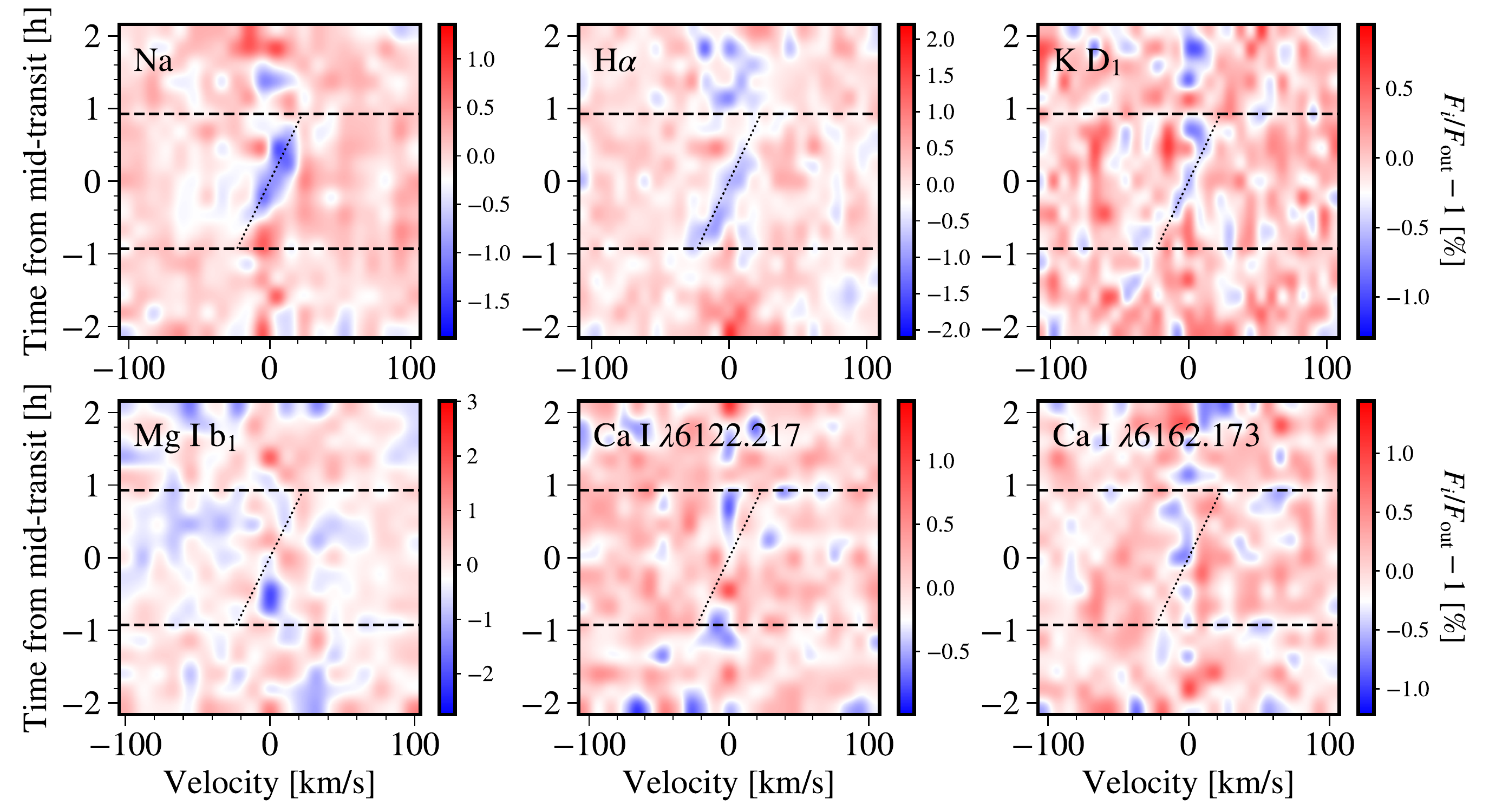}
\caption{Phase-resolved transmission spectrum at the Na, H$\alpha$, K D$_1$ lines (top row) and three of the control lines (bottom row). Signatures arising from the center-to-limb variation and Rossiter-McLaughlin effects have been corrected. Three nights have been combined and rebinned to the phase and velocity grid. Dashed lines mark the first and fourth contacts of the transit. Slanted dotted lines indicate the radial velocity shift induced by the planet orbital motion at the expected radial velocity semi-amplitude $K_\mathrm{p}=167$~km\,s$^{-1}$. The top row shows an excess absorption (traced in blue color) with a velocity shift during the transit, which agrees with the expected planet orbital motion, while the bottom row does not exhibit any significant excess absorption with a regular velocity shift.}
\label{fig:planetrv}
\end{figure*}

\begin{figure*}
\centering
\includegraphics[width=0.7\linewidth]{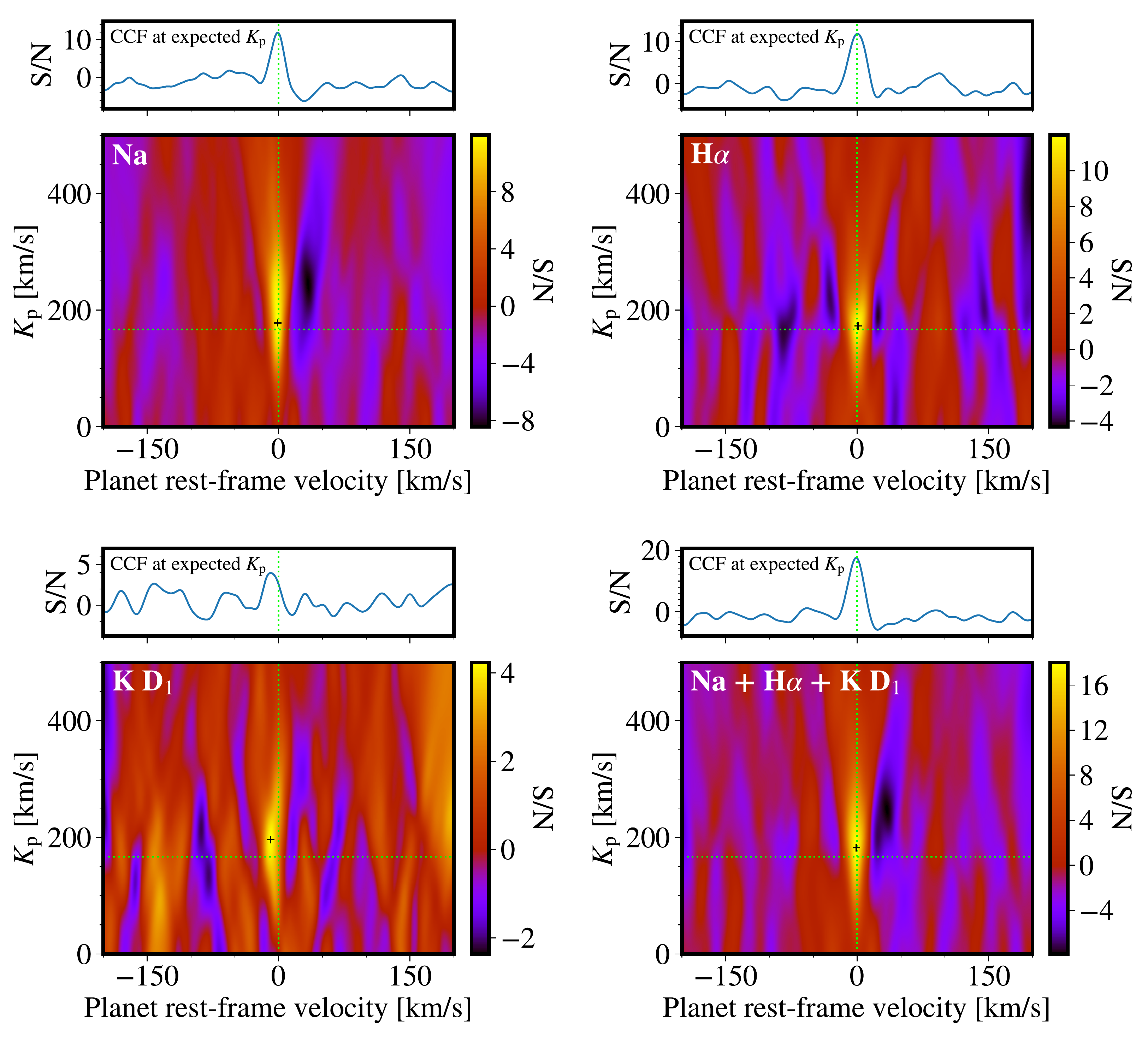}
\caption{Combined cross-correlation function (CCF) calculated at pairs of $K_\mathrm{p}$ and velocity for Na (top left), H$\alpha$ (top right), K D$_1$ (bottom left), and Na+H$\alpha$+K~D$_1$ combined (bottom right). The CCF has been normalized and expressed in the form of S/N. The plus sign marks the maximum S/N location. The dotted lines refer to $K_\mathrm{p}=167$~km\,s$^{-1}$ and zero velocity. The sub-panel above the $K_\mathrm{p}$ map shows the normalized CCF at $K_\mathrm{p}=167$~km\,s$^{-1}$.}
\label{fig:ccmap}
\end{figure*}

\subsection{Empirical Monte Carlo simulation for absorption depth measurements}
\label{sec:emc}

We employed the empirical Monte Carlo (EMC) approach \citep[e.g.,][]{2008ApJ...673L..87R} to assess the systematic effects and to ensure that the excess absorption is only detected when the planet is in transit. We performed the EMC simulation on data from the passbands that achieve best S/N, except for the control lines, where we simply chose the 0.4~$\AA$ band. The calculation was done on three simulated data sets. For the ``in-out'' scenario, we randomly selected a subset from the real in-transit data as the simulated ``in'', with the selecting fraction being no less than half, and a subset from the real out-of-transit data as the simulated ``out'', with its number being proportional to the real out-to-in ratio. For the ``in-in'' scenario, we randomly divided the real in-transit data into two subsets, and the two subsets were assumed as simulated ``in'' and ``out'', with their number ratio being the same as the real in-to-out ratio. The ``out-out'' scenario is similar to ``in-in'', but changed its source as the real out-of-transit data. The absorption depth is measured on the simulated data sets in the same way as described in Sect.~\ref{sec:detection}. We take the median of ``in-out'' as the best-fit value and the scaled standard deviation of ``out-out'' as the uncertainty. The scaling factor is adopted as the square root of the ratio of the number of out-of-transit spectra to the number of all the spectra.

Fig.~\ref{fig:emc} presents the resulting posterior distributions from the EMC simulation. The excess absorption signal is only detected in the ``in-out'' scenario of the Na, H$\alpha$, and K D$_1$ lines. The distributions of ``in-in'' and ``out-out'' are centered at zero. We obtained $0.88\pm 0.16$~\% (2$\times$0.4~$\AA$ Na band), $0.70\pm 0.25$~\% (0.35~$\AA$ H$\alpha$ band), and $0.30\pm 0.09$~\% (0.35~$\AA$ K D$_1$ band), which are consistent with what we obtained on the real data, as derived in Sect.~\ref{sec:binned_ad}.

\subsection{Confirmation of planetary origin for Na, H$\alpha$, K D$_1$}

Phase-resolved high-resolution transmission spectrum can trace the planet motion in radial velocity and thus confirm the origin of the observed excess absorption \citep[e.g.,][]{2010Natur.465.1049S,2019A&A...628A...9C}. In our case, the stellar line center, where the planet excess absorption is located (expected to shift between $-$23~km\,s$^{-1}$ and $+$23~km\,s$^{-1}$, i.e., $\sim$$-$0.5~$\AA$ to $\sim$$+$0.5~$\AA$), is too noisy to exhibit a significant planet trace. Nevertheless, Fig.~\ref{fig:planetrv} presents a Gaussian-convolved plot of the phase-solved transmission spectrum at the Na, H$\alpha$, and K D$_1$ lines, and three stellar activity indicator control lines. The velocity shifts of the Na, H$\alpha$, and K D$_1$ lines are fully consistent with that induced by the planet orbital motion during the transit. However, excess absorption still exists after the transit, which might be due to even lower S/N because those phase were observed at much higher airmass. 

We further derived the planet RV semi-amplitude $K_\mathrm{p}$ from the Na, H$\alpha$, and K D$_1$ lines to quantitatively confirm whether or not the signal is planet-correlated. Given the low S/N, it is difficult to directly fit for $K_\mathrm{p}$ on the phase-resolved transmission spectrum. Here we employed the cross-correlation technique to combine in-transit signals to derive $K_\mathrm{p}$. The cross-correlation technique has been widely used to enhance the absorption signal by taking advantage of line forest if the absorption lines are weak \citep[e.g.,][]{2010Natur.465.1049S,2012Natur.486..502B}. We used the best-fit Gaussian function in Sect.~\ref{sec:detection} as the cross-correlation model template, and shifted it from $-$200 to $+$200~km\,s$^{-1}$ in steps of 1~km\,s$^{-1}$. This velocity shift is denoted as $\Delta v_\mathrm{p}$. The cross-correlation function (CCF) was calculated for each exposure at each $\Delta v_\mathrm{p}$. For a given $K_\mathrm{p}$ in the grid of 0--500~km\,s$^{-1}$ (in steps of 1~km\,s$^{-1}$), the in-transit CCF were shifted to planet-rest frame and combined. We normalized the $K_\mathrm{p}$-$\Delta v_\mathrm{p}$ map by the standard deviation of CCF within the velocity ranges of $-$200 to $-$100 km\,s$^{-1}$ and $+$100 to $+$200 km\,s$^{-1}$. The resulting $K_\mathrm{p}$-$\Delta v_\mathrm{p}$ S/N map for the three lines and their combination is shown in Fig.~\ref{fig:ccmap}.

The peak S/N occurred at ($K_\mathrm{p}$, $\Delta v_\mathrm{p}$) = ($178^{+76}_{-36}$, $-1^{+3}_{-3}$) km\,s$^{-1}$ for Na, at ($173^{+20}_{-23}$, $+1^{+4}_{-5}$) km\,s$^{-1}$ for H$\alpha$, and at ($196^{+47}_{-63}$, $-9^{+7}_{-6}$) km\,s$^{-1}$ for K D$_1$. The associated uncertainties correspond to the 1$\sigma$ contour around the peak \citep[e.g.][]{2018A&A...615A..16B,2019MNRAS.482.4422C}. Since we did not observe any significant blueshift or redshift when fitting Gaussian function to the line profile (see Sect.~\ref{sec:detection}), we decided to adopt $K_\mathrm{p}$ at zero velocity and derived 178$^{+75}_{-35}$~km\,s$^{-1}$ for Na, 172$^{+20}_{-24}$~km\,s$^{-1}$ for H$\alpha$, and 255$^{+75}_{-155}$~km\,s$^{-1}$ for K D$_1$. Na and K D$_1$ have less constrained $K_\mathrm{p}$ due to their S/N being lower than H$\alpha$ at the line center. All three values are consistent with the expected value induced by the planet orbital motion: $K_\mathrm{p}=167\pm 11$~km\,s$^{-1}$. Consequently, the detected excess absorption at the Na, H$\alpha$, and K D$_1$ lines are very likely originated in the atmosphere of the planet.

\section{Discussion}
\label{sec:discuss}

\begin{figure*}
\centering
\includegraphics[width=\linewidth]{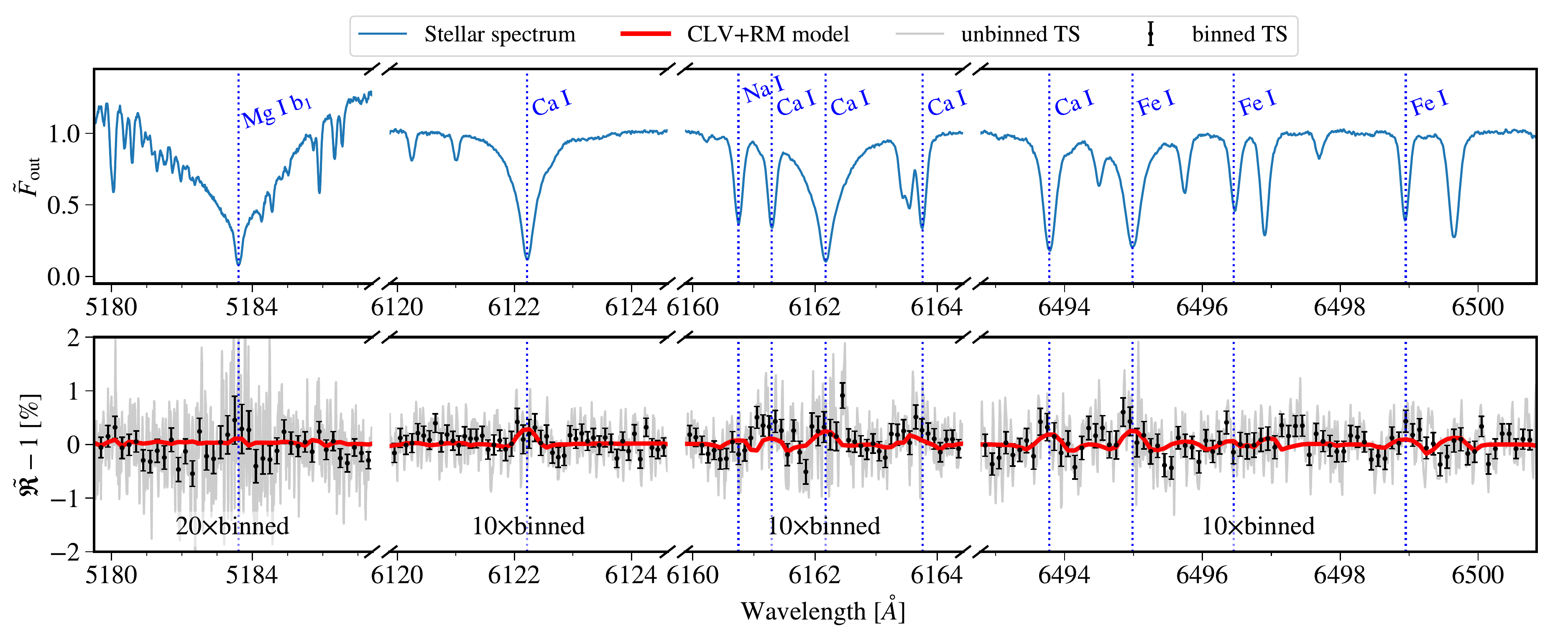}
\caption{Top panel shows the master-out stellar spectrum at the stellar activity indicator control lines. Bottom panel shows the corresponding ``transmission spectrum'' (TS), where the center-to-limb variation and Rossiter-McLaughlin (CLV+RM) effects have not been corrected. The gray line and black circles refer to the unbinned and binned TS, respectively. The red line shows the CLV+RM model.}
\label{fig:activity_indicators}
\end{figure*}

\begin{figure*}
\centering
\includegraphics[width=\linewidth]{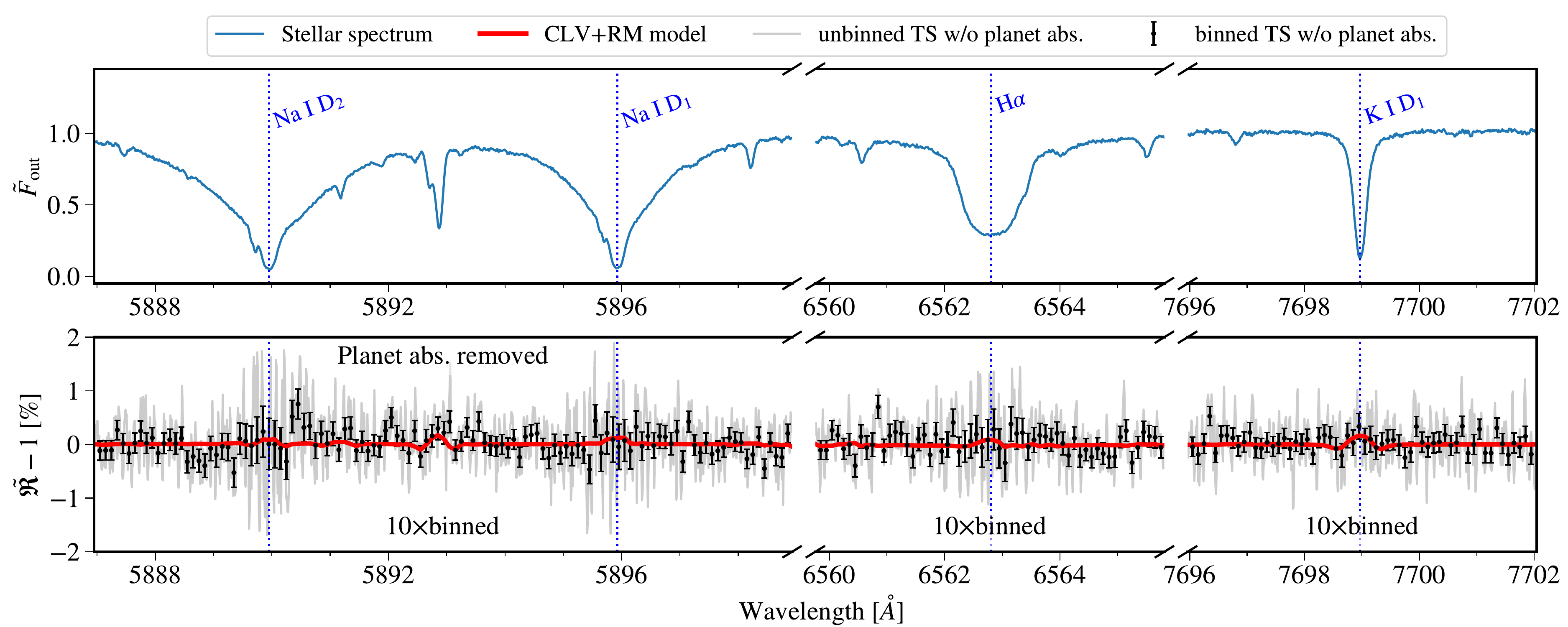}
\caption{Top panel shows the master-out stellar spectrum at the Na doublet, H$\alpha$, and K D$_1$ lines. Bottom panel shows the corresponding ``transmission spectrum'' (TS), where the center-to-limb variation and Rossiter-McLaughlin (CLV+RM) effects have not been corrected and the best-fit planetary absorption signals have been removed. The gray line and black circles refer to the unbinned and binned TS, respectively. The red line shows the CLV+RM model.}
\label{fig:clv_rm_only}
\end{figure*}

\subsection{Stellar activity indicator lines as the control sample}
\label{sec:control}

Both the H$\alpha$ line \citep[e.g.,][]{1991A&A...251..199P,2007A&A...469..309C} and the \ion{Na}{I} D doublet \citep[e.g.,][]{1997A&A...322..266A,2007MNRAS.378.1007D} have been widely used as the stellar chromospheric activity indicators for low-mass stars and very active stars. \citet{2018AJ....156..189C} performed simulations of stellar active regions to assess the impact on the high-resolution transmission spectrum, and found that strong facular emission and large coverage fractions can contribute non-negligible contaminations at the H$\alpha$ and \ion{Na}{I} D lines. Given that WASP-52 is a moderately active K2 dwarf, we used the \texttt{ACTIN} code \citep{2011A&A...534A..30G,2018JOSS....3..667G} to measure the \ion{Ca}{II}\,H\&K S-index, the H$\alpha$ index, and  the \ion{Na}{I} index for every exposure. We obtained range of values 0.476--0.506 for the \ion{Ca}{II}\,H\&K S-index, 0.240--0.245 for the H$\alpha$ index, and 0.154--0.162 for the \ion{Na}{I} index, respectively. We did not find any significant correlation between the \ion{Ca}{II}\,H\&K S-index and the H$\alpha$ index, nor between the \ion{Ca}{II}\,H\&K S-index and the \ion{Na}{I} index. 

We then examined a few other stellar activity indicator lines as the control experiment. We collected the stellar activity indicator lines from \citet{2010MNRAS.403.2157H} and \citet{2019MNRAS.490L..86Y}, including: \ion{Mg}{I} b$_1$ 5183.604$\AA$, \ion{Ca}{I} 6122.217$\AA$, \ion{Na}{I} 6160.747$\AA$, \ion{Ca}{I} 6161.29$\AA$, \ion{Ca}{I} 6162.173$\AA$, \ion{Ca}{I} 6163.76$\AA$, \ion{Ca}{I} 6496.456$\AA$, \ion{Fe}{I} 6494.985$\AA$, \ion{Fe}{I} 6496.456$\AA$, \ion{Fe}{I} 6498.950$\AA$. Among these lines, \ion{Mg}{I} b$_1$ 5183.604$\AA$, \ion{Ca}{I} 6122.217$\AA$, and \ion{Ca}{I} 6162.173$\AA$ have served as the control lines in past studies \citep{2017A&A...602A..36W,2019AJ....158..120Z}.

We performed the binned absorption depth analysis (see Fig.~\ref{fig:adgrowth}) and the EMC simulation analysis (see Fig.~\ref{fig:emc}) on these control lines in the same way as with the planetary lines. We found that none of these control lines present excess absorption depth values significantly offset from zero. Only the \ion{Na}{I} 6160.747$\AA$ line exhibits an absorption depth growth curve similar to the planetary originated lines, while some others (e.g., \ion{Mg}{I} b$_1$ 5183.604$\AA$ and \ion{Ca}{I} 6161.29$\AA$) present ``negative'' absorption depths, which might be due to the imperfect correction of the CLV+RM effects. We note that we only performed a nominal CLV+RM correction assuming LTE and a planetary radius of 1\,$R_\mathrm{p}$ due to limited S/N. A further visual inspection of the ``transmission spectrum'' at the \ion{Na}{I} 6160.747$\AA$ line does not reveal a well shaped line profile, which is probably just noise. 

Fig.~\ref{fig:activity_indicators} presents the derived transmission spectrum zoomed at these control lines. For comparison, Fig.~\ref{fig:clv_rm_only} shows the planetary line ``transmission spectrum'' after removing the best-fit planetary absorption. In both cases, the CLV+RM effects have not been corrected. The control line ``transmission spectrum'' does not show any significant planetary absorption, but it does show evidences of the CLV+RM effects as indicated by the nominal model. On the other hand, Fig.~\ref{fig:clv_rm_only} confirms that the planetary absorption at the Na, H$\alpha$, and K D$_1$ lines are not significantly affected by the CLV+RM effects. Furthermore, as shown in Fig.~\ref{fig:planetrv}, these control lines do not exhibit the velocity shift trace that is consistent with the planet orbital motion during the transit.

\subsection{Impact of the CLV and RM effects on transmission spectrum}
\label{sec:control}

\begin{figure}
\centering
\includegraphics[width=\linewidth]{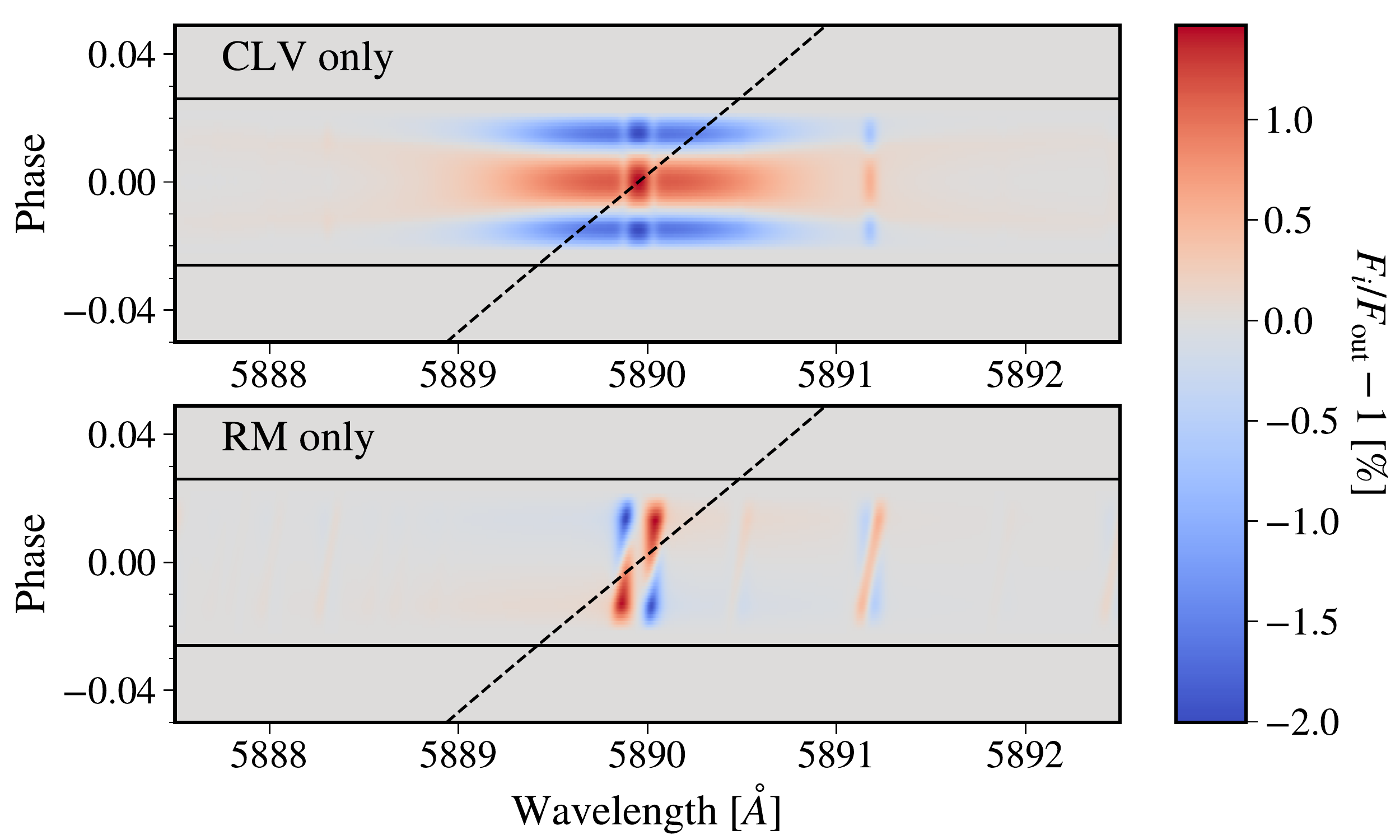}
\caption{Individual contributions of the center-to-limb variation (CLV) and Rossiter-McLaughlin (RM) effects at the Na D$_2$ line for WASP-52b. The top and bottom panels show the CLV-only and RM-only models, respectively. The horizontal lines mark the first and fourth contacts of the transit. The dashed line marks the trace of planet orbital motion.}
\label{fig:clvrm_sep}
\end{figure}

\begin{figure}
\centering
\includegraphics[width=\linewidth]{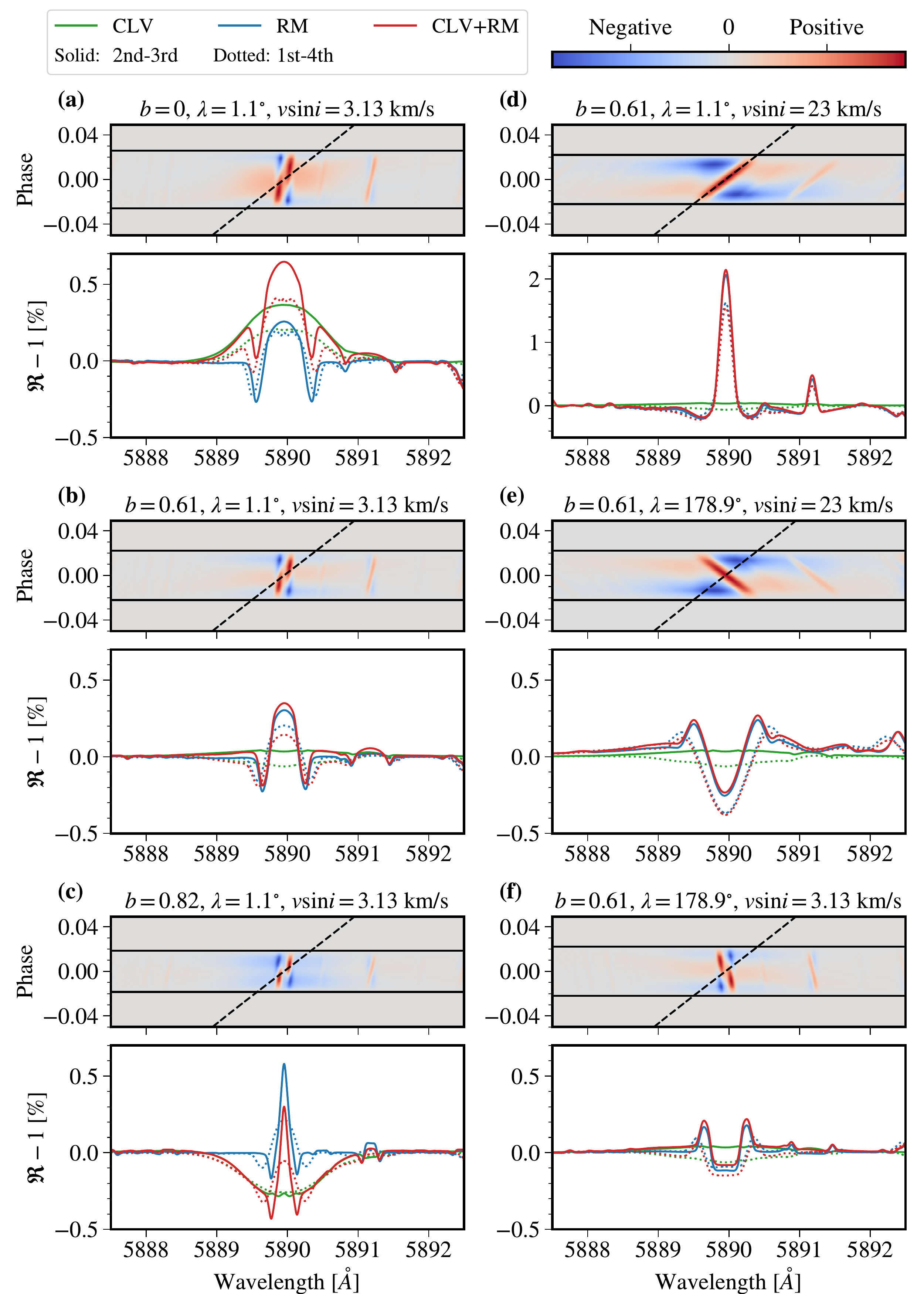}
\caption{Artifact transmission spectrum with the center-to-limb variation (CLV) and Rossiter-McLaughlin (RM) effects. In each panel, the top sub-panel shows the phase-resolved CLV+RM combined model in the stellar-rest frame, where the first and fourth contacts of the transit are marked by horizontal lines and the planet orbital motion is traced by the dashed line. The bottom sub-panel shows the in-transit combined (between the first and fourth contacts, dotted line) and the fully in-transit combined (between the second and third contacts, solid line) ``transmission spectra'' in the planet-rest frame. The CLV-only, RM-only, and CLV+RM combined models are shown in green, blue, and red colors, respectively. Panels (a), (b), (c) present experiments with different impact parameters, where (b) refers to the real case of WASP-52b. Panels (d), (e), (f) adopt the same impact parameters as WASP-52b, but employing a different $v\sin i$ in (d) and (e), or a different $\lambda$ in (e) and (f). }
\label{fig:clvrm_experiment}
\end{figure}

\citet{CasasayasBarris2020} has shown that a non-correction of the CLV+RM effects would potentially introduce artifacts distorting the transmission spectrum or result in false positive detections of atomic species in planetary atmospheres. In Fig.~\ref{fig:clvrm_sep}, we present the individual contributions of the CLV-only and RM-only models for WASP-52b. The CLV effect does not show any velocity shift feature in the phase-resolved transmission spectrum matrix in the stellar-rest frame. In contrast, the RM effect introduces an X-shaped feature in the velocity domain, one arm with positive values in a similar direction to the planetary orbital trace while the other arm with negative values perpendicular to the trace. The combined CLV+RM effects in the planet-rest frame resemble bump-like features (see Fig.~\ref{fig:activity_indicators} and Fig.~\ref{fig:clv_rm_only}).

Although it does not affect our detections of the Na, H$\alpha$, and K D$_1$ lines even without the CLV+RM correction (see Fig.~\ref{fig:clv_rm_only}), it would potentially cancel out the planetary absorption if the signal were weak. Therefore, we performed experiments to investigate how the CLV+RM effects would impact the transmission spectrum if the planetary system has a different impact parameter, a different projected stellar rotation velocity, or a different projected spin-orbit angle. 

First, we started with the same parameters as the WASP-52 system (see Table \ref{tab:param}), where the impact parameter is $b=0.61$. We varied the impact parameter to $b=0$ and $b=0.82$ by changing the orbital inclination to $i=90^{\circ}$ and $i=83.4^{\circ}$, so as to investigate the dependence of the CLV+RM effects on the impact parameter. Then, we went back to the WASP-52 system parameters and changed the projected stellar rotation velocity to $v\sin i=23$~km\,s$^{-1}$. Finally, we flipped the projected spin-orbit angle to $\lambda=178.9^{\circ}$ to investigate the retrograde orbit.

We present the results of the experiments in Fig.~\ref{fig:clvrm_experiment}, using the Na D$_2$ line for illustration. When increasing the impact parameter, the transit duration shrinks, the RM effect becomes sharper, while the CLV effect first appears as a bump-like feature (see Fig.~\ref{fig:clvrm_experiment}a), and then decreases to an artifact absorption feature (see Fig.~\ref{fig:clvrm_experiment}c). The actual case of WASP-52b coincidently has a very minimized impact of the CLV effect (see Fig.~\ref{fig:clvrm_experiment}b). This changing trend is similar to what is observed in the extensive CLV-only experiments of \citet{2017A&A...603A..73Y} (see their Fig.~9). On the other hand, when the projected stellar rotation velocity is approximately $v\sin i=23$~km\,s$^{-1}$, the positive arm of the X-shaped feature is aligned  to the planetary orbital trace. Consequently, the RM effect is maximized (see Fig.~\ref{fig:clvrm_experiment}d). If the projected spin-orbit angle is mirrored to a retrograde orbit, the negative arm of the X-shaped feature is approximately aligned to the planetary trace (see Fig.~\ref{fig:clvrm_experiment}e), and the resulting transmission spectrum would cause a false positive detection if no CLV+RM correction has been performed. The artifact shape looks less like an absorption feature when the projected rotation velocity decreases back to that of WASP-52's (see Fig.~\ref{fig:clvrm_experiment}f). 

As a brief summary, the CLV+RM effects can introduce a false positive detection or cancel out a real signal. We also note that combining the spectra of in-transit (between the first and fourth contacts) or fully in-transit (between the second and third contacts) could also result in different levels of impact. Our experiments have been conducted in a uniform and fine sampled grid (0.003~$\AA$ in wavelength and 0.001 in phase). However, a real observation would sample the phase asymmetric to the transit center with overhead gaps. The exposure time could be too long to sample the phase well. The implementation of weighted average would add another complexity. A lower spectral resolution than the wavelength sampling in our experiments could also smear out sharp features, resulting in Gaussian like features. Therefore, the CLV+RM effects have to be examined on each observation before any claims of detection or non-detection of certain species \citep{CasasayasBarris2020}.

\subsection{The atmosphere of WASP-52b}
\label{sec:w52atm}

WASP-52b has been studied at low-resolution by transmission spectroscopy with different instruments, covering from optical to infrared wavelengths \citep{2017A&A...600L..11C,2017MNRAS.470..742L,2018AJ....156..122M,2018AJ....156..124B,2018AJ....156..298A}. While the 1.4~$\mu$m water absorption signature has been detected in the near infrared \citep{2018AJ....156..124B}, the lack of pressure-broadened Na and K absorption line wings in the optical wavelengths implies the presence of a cloud deck at high altitude. The moderate stellar activity complicates the interpretation of the transmission spectra acquired at different epochs, which might account for the overall offset between the flat transmission spectra obtained by GTC/OSIRIS \citep{2017A&A...600L..11C} and HST/STIS \citep{2018AJ....156..298A}. While \citet{2017A&A...600L..11C} reported a 3.6$\sigma$ detection of Na and a 2.2$\sigma$ detection of K above the cloud deck, \citet{2018AJ....156..298A} can only confirm the Na detection at 2.3$\sigma$ without any evidence of K. The lower significance of the HST Na detection and K non-detection might be related to the larger bin size used to derive the transit depth, where it is 35~$\AA$ v.s. 16~$\AA$ for Na and 75~$\AA$ v.s. 16~$\AA$ for K. \citet{2020MNRAS.491.5361B} performed the atmospheric spectral retrievals on WASP-52b's full optical to infrared (0.3--5~$\mu$m, HST/STIS+HST/WFC3+Spitzer/IRAC) low-resolution transmission spectrum, allowing stellar activity to account for the offset between the HST/STIS and HST/WFC3 measurements, and found a 0.1--10$\times$ solar metallicity, a sub-solar C/O ratio, and a solar-like water abundance for the planetary atmosphere. However, the abundances of Na and K are still not well constrained.

Our new ESPRESSO transmission spectrum is able to not only confirm the presence of Na and K in the upper atmosphere of WASP-52b, but also reveal a new detection of neutral hydrogen in the form of H$\alpha$, demonstrating the great power of high-resolution transmission spectroscopy, more so in the case of cloudy atmospheres. Assuming an unresolved Na doublet, the ESPRESSO Na detection convolved to the resolution of the GTC/OSIRIS data shows a consistent excess absorption depth at the Na line ($\sim$0.385\% for ESPRESSO v.s. $\sim$0.378\% for GTC/OSIRIS in a 16~$\AA$ band). This makes WASP-52b one of the very few planets for which the same atomic species has been confirmed by both low- and high-resolution observations, e.g., the Na doublet seen in HD 189733b \citep{2012MNRAS.422.2477H,2015A&A...577A..62W}, the \ion{He}{I} 1083~nm triplet seen in WASP-107b \citep{2018Natur.557...68S,2019A&A...623A..58A} and HAT-P-11b \citep{2018ApJ...868L..34M,2018Sci...362.1384A}.

The resolved line profiles of the three species could be linked to WASP-52b's atmospheric properties. According to the measured line contrast, the line cores of Na D$_1$ and D$_2$ probe the atmospheric layer at $\sim$1.19--1.23~$R_\mathrm{p}$, the line core of K D$_1$ probes $\sim$1.09~$R_\mathrm{p}$, and the line core of H$\alpha$ probes $\sim$1.16~$R_\mathrm{p}$. All of these line cores are well below the effective Roche lobe radius, derived to be $1.72\pm 0.05$~$R_\mathrm{p}$ based on the equation (2.5) of \citet{2010EAS....41..429E}. Consequently, we are not observing any atmospheric escaping signature at these tracer lines. The offset of the line center in the planet-rest frame is in general attributed to planetary wind in the upper atmosphere. The line center of Na D$_2$ shows a marginal net blueshift  ($-2.6\pm 1.2$~km\,s$^{-1}$). However, the overall weighted average offset of all the lines, $-0.2\pm 0.6$~km\,s$^{-1}$, is well consistent with zero, indicating that we are not observing any planetary wind. 

Given that the difference in the transit radius at the line cores of Na D$_1$ and D$_2$ corresponds to $\ln(2)$ times of the atmospheric scale height \citep{2017ApJ...851..150H}, we can estimate a temperature of $\sim$4630~K if it is dominated by atomic hydrogen ($\mu=1.3$). For H$\alpha$, we have measured a line FWHM of 15.4~km\,s$^{-1}$, i.e., $\sigma=\mathrm{FWHM}/2.355=6.5$~km\,s$^{-1}$. Assuming that H$\alpha$ originates in a similar layer to the Na doublet and that its line width is set by thermal broadening and the maximum optical depth at line center, based on the equations (3) and (4) of \citet{2017ApJ...851..150H}, we can derive an optical depth of $\tau_0=\exp[\sigma^2m_p/(2k_\mathrm{B}T)]\sim$1.7 and a number density of $n_{2l}=35\times 10^4\tau_0\sim 500$~cm$^{-3}$ for the hydrogen 2$l$ state. 

In addition to WASP-52b, five other planets have simultaneous detections of Na and H$\alpha$, including HD 189733b \citep{2015A&A...577A..62W,2012ApJ...751...86J,2015ApJ...810...13C,2016AJ....152...20C,2017ApJ...851..150H}, KELT-20b \citep{2019A&A...628A...9C}, KELT-9b \citep{2018NatAs...2..714Y,2019AJ....157...69C}, WASP-12b \citep{2018AJ....156..154J}. and WASP-121b \citep{2020arXiv200107196C}. However, WASP-52b is currently the only planet whose H$\alpha$ absorption is shallower than that of Na. Furthermore, WASP-52b is the second non-ultra-hot Jupiter to exhibit H$\alpha$ absorption, only second to HD 189733b, which also orbits an active K dwarf and has a similar equilibrium temperature ($\sim$1200~K). The detection of H$\alpha$ indicates that the probed atmospheric layer must be very hot. Given their relatively low equilibrium temperature, high XUV flux is required to heat the upper atmosphere, which is likely correlated with stellar activity. Since we have combined three transits, it is possible that the H$\alpha$ transit depth is variable during our three transits depending on the Ly$\alpha$ intensity \citep{2017ApJ...851..150H}. The combined results might have averaged down the H$\alpha$ transit depth, making it shallower than the relatively constant Na transit depth. Unfortunately, our current observations, three transits combined, are already limited by S/N. Future follow-up observations, designed in groups of transits to increase S/N and each group at different stellar activity levels, could help reveal the cause of shallower H$\alpha$ absorption. This non-ultra-hot Jupiter population could play a crucial role in understanding the star-planet interaction in the era of extremely large telescopes.

\section{Conclusions}\label{sec:conclusions}

We observed three transits of the hot Jupiter WASP-52b using the ultra-stable high-resolution spectrograph ESPRESSO at the VLT. We collected a total of 58 spectra, 28 of which were acquired during out-of-transit. We corrected telluric contamination, stellar center-to-limb variation and Rossiter-McLaughlin effect in the observed stellar spectra, and derived a residual spectrum matrix that only contains astrophysical signal of planetary origin. With the combined transmission spectrum in the planet-rest frame, we spectrally resolved the excess absorption at Na doublet, H$\alpha$, and K D$_1$, but did not detect any excess at other lines. We fitted Gaussians to the line profiles of detected lines, but we did not find any significant offset of the line center, indicating that we have not observed any evidence for planetary wind. The measured line contrast is $\sim$1.09\% for Na D$_1$,  $\sim$1.31\% for Na D$_2$,  $\sim$0.86\% for H$\alpha$, and $\sim$0.46\% for K D$_1$, respectively.

We performed the empirical Monte Carlo simulation and confirmed that the excess absorption at Na, H$\alpha$, and K D$_1$ can only be reproduced if we compare the in-transit to the out-of-transit data. We further confirm that the excess absorption comes from planet atmosphere using the cross-correlation technique, given that the signal's RV shift is fully consistent with the planet orbital motion. We used stellar activity indicator lines to conduct the control experiment, and did not repeat any of the detections as obtained at the Na, H$\alpha$, and K D$_1$ lines, confirming that the potential stellar activity does not affect our results.

\begin{acknowledgements}
    G. C. acknowledges the support by the Natural Science Foundation of Jiangsu Province (Grant No. BK20190110), the National Natural Science Foundation of China (Grant No. 11503088, 11573073, 11573075), and the Minor Planet Foundation of the Purple Mountain Observatory. This work is partly financed by the Spanish Ministry of Economics and Competitiveness through project ESP2016-80435-C2-2-R. F. Y. acknowledges the support of the DFG priority program SPP 1992 ``Exploring the Diversity of Extrasolar Planets (RE 1664/16-1)''. This work has been partly carried out within the frame of the National Centre for Competence in Research ``PlanetS'' supported by the Swiss National Science Foundation (SNSF). H.M.C. and R.A. acknowledge the financial support of the SNSF. 
    This work has made use of the VALD database, operated at Uppsala University, the Institute of Astronomy RAS in Moscow, and the University of Vienna.
    This research has made use of Matplotlib \citep{2007CSE.....9...90H}, the VizieR catalog access tool, CDS, Strasbourg, France \citep{2000A&AS..143...23O}, and TEPCat \citep{2011MNRAS.417.2166S}. 
    The authors thank the anonymous referee for constructive comments that have improved the manuscript.
\end{acknowledgements}

\bibliographystyle{aa} 
\bibliography{ref_db.bib} 

\begin{thebibliography}{104}
\expandafter\ifx\csname natexlab\endcsname\relax\def\natexlab#1{#1}\fi

\bibitem[{{Akinsanmi} {et~al.}(2018){Akinsanmi}, {Oshagh}, {Santos}, \&
  {Barros}}]{2018A&A...609A..21A}
{Akinsanmi}, B., {Oshagh}, M., {Santos}, N.~C., \& {Barros}, S.~C.~C. 2018,
  \aap, 609, A21

\bibitem[{{Alam} {et~al.}(2018){Alam}, {Nikolov}, {L{\'o}pez-Morales}, {Sing},
  {Goyal}, {Henry}, {Sanz-Forcada}, {Williamson}, {Evans}, {Wakeford}, {Bruno},
  {Ballester}, {Stevenson}, {Lewis}, {Barstow}, {Bourrier}, {Buchhave},
  {Ehrenreich}, \& {Garc{\'{\i}}a Mu{\~n}oz}}]{2018AJ....156..298A}
{Alam}, M.~K., {Nikolov}, N., {L{\'o}pez-Morales}, M., {et~al.} 2018, \aj, 156,
  298

\bibitem[{{Albrecht} {et~al.}(2012){Albrecht}, {Winn}, {Johnson}, {Howard},
  {Marcy}, {Butler}, {Arriagada}, {Crane}, {Shectman}, {Thompson}, {Hirano},
  {Bakos}, \& {Hartman}}]{2012ApJ...757...18A}
{Albrecht}, S., {Winn}, J.~N., {Johnson}, J.~A., {et~al.} 2012, \apj, 757, 18

\bibitem[{{Allart} {et~al.}(2019){Allart}, {Bourrier}, {Lovis}, {Ehrenreich},
  {Aceituno}, {Guijarro}, {Pepe}, {Sing}, {Spake}, \&
  {Wyttenbach}}]{2019A&A...623A..58A}
{Allart}, R., {Bourrier}, V., {Lovis}, C., {et~al.} 2019, \aap, 623, A58

\bibitem[{{Allart} {et~al.}(2018){Allart}, {Bourrier}, {Lovis}, {Ehrenreich},
  {Spake}, {Wyttenbach}, {Pino}, {Pepe}, {Sing}, \& {Lecavelier des
  Etangs}}]{2018Sci...362.1384A}
{Allart}, R., {Bourrier}, V., {Lovis}, C., {et~al.} 2018, Science, 362, 1384

\bibitem[{{Allart} {et~al.}(2017){Allart}, {Lovis}, {Pino}, {Wyttenbach},
  {Ehrenreich}, \& {Pepe}}]{2017A&A...606A.144A}
{Allart}, R., {Lovis}, C., {Pino}, L., {et~al.} 2017, \aap, 606, A144

\bibitem[{{Alonso-Floriano} {et~al.}(2019){Alonso-Floriano}, {Snellen},
  {Czesla}, {Bauer}, {Salz}, {Lamp{\'o}n}, {Lara}, {Nagel},
  {L{\'o}pez-Puertas}, {Nortmann}, {S{\'a}nchez-L{\'o}pez}, {Sanz-Forcada},
  {Caballero}, {Reiners}, {Ribas}, {Quirrenbach}, {Amado}, {Aceituno},
  {Anglada-Escud{\'e}}, {B{\'e}jar}, {Brinkm{\"o}ller}, {Hatzes}, {Henning},
  {Kaminski}, {K{\"u}rster}, {Labarga}, {Montes}, {Pall{\'e}}, {Schmitt}, \&
  {Zapatero Osorio}}]{2019A&A...629A.110A}
{Alonso-Floriano}, F.~J., {Snellen}, I.~A.~G., {Czesla}, S., {et~al.} 2019,
  \aap, 629, A110

\bibitem[{{Andretta} {et~al.}(1997){Andretta}, {Doyle}, \&
  {Byrne}}]{1997A&A...322..266A}
{Andretta}, V., {Doyle}, J.~G., \& {Byrne}, P.~B. 1997, \aap, 322, 266

\bibitem[{{Bourrier} {et~al.}(2017){Bourrier}, {Cegla}, {Lovis}, \&
  {Wyttenbach}}]{2017A&A...599A..33B}
{Bourrier}, V., {Cegla}, H.~M., {Lovis}, C., \& {Wyttenbach}, A. 2017, \aap,
  599, A33

\bibitem[{{Bourrier} {et~al.}(2020){Bourrier}, {Ehrenreich}, {Lendl},
  {Cretignier}, {Allart}, {Dumusque}, {Cegla}, {Suarez-Mascareno},
  {Wyttenbach}, {Hoeijmakers}, {Melo}, {Kuntzer}, {Astudillo-Defru}, {Giles},
  {Heng}, {Kitzmann}, {Lavie}, {Lovis}, {Murgas}, {Nascimbeni}, {Pepe}, {Pino},
  {Segransan}, \& {Udry}}]{2020arXiv200106836B}
{Bourrier}, V., {Ehrenreich}, D., {Lendl}, M., {et~al.} 2020, arXiv e-prints,
  arXiv:2001.06836

\bibitem[{{Bourrier} {et~al.}(2018){Bourrier}, {Lovis}, {Beust}, {Ehrenreich},
  {Henry}, {Astudillo-Defru}, {Allart}, {Bonfils}, {S{\'e}gransan}, {Delfosse},
  {Cegla}, {Wyttenbach}, {Heng}, {Lavie}, \& {Pepe}}]{2018Natur.553..477B}
{Bourrier}, V., {Lovis}, C., {Beust}, H., {et~al.} 2018, \nat, 553, 477

\bibitem[{{Brahm} {et~al.}(2017){Brahm}, {Jord{\'a}n}, {Hartman}, \&
  {Bakos}}]{2017MNRAS.467..971B}
{Brahm}, R., {Jord{\'a}n}, A., {Hartman}, J., \& {Bakos}, G. 2017, \mnras, 467,
  971

\bibitem[{{Brogi} {et~al.}(2018){Brogi}, {Giacobbe}, {Guilluy}, {de Kok},
  {Sozzetti}, {Mancini}, \& {Bonomo}}]{2018A&A...615A..16B}
{Brogi}, M., {Giacobbe}, P., {Guilluy}, G., {et~al.} 2018, \aap, 615, A16

\bibitem[{{Brogi} {et~al.}(2012){Brogi}, {Snellen}, {de Kok}, {Albrecht},
  {Birkby}, \& {de Mooij}}]{2012Natur.486..502B}
{Brogi}, M., {Snellen}, I.~A.~G., {de Kok}, R.~J., {et~al.} 2012, \nat, 486,
  502

\bibitem[{{Brown} {et~al.}(2017){Brown}, {Triaud}, {Doyle}, {Gillon}, {Lendl},
  {Anderson}, {Collier Cameron}, {H{\'e}brard}, {Hellier}, {Lovis}, {Maxted},
  {Pepe}, {Pollacco}, {Queloz}, \& {Smalley}}]{2017MNRAS.464..810B}
{Brown}, D.~J.~A., {Triaud}, A.~H.~M.~J., {Doyle}, A.~P., {et~al.} 2017,
  \mnras, 464, 810

\bibitem[{{Bruno} {et~al.}(2020){Bruno}, {Lewis}, {Alam}, {L{\'o}pez-Morales},
  {Barstow}, {Wakeford}, {Sing}, {Henry}, {Ballester}, {Bourrier}, {Buchhave},
  {Cohen}, {Mikal-Evans}, {Garc{\'\i}a Mu{\~n}oz}, {Lavvas}, \&
  {Sanz-Forcada}}]{2020MNRAS.491.5361B}
{Bruno}, G., {Lewis}, N.~K., {Alam}, M.~K., {et~al.} 2020, \mnras, 491, 5361

\bibitem[{{Bruno} {et~al.}(2018){Bruno}, {Lewis}, {Stevenson}, {Filippazzo},
  {Hill}, {Fraine}, {Wakeford}, {Deming}, {L{\'o}pez-Morales}, \&
  {Alam}}]{2018AJ....156..124B}
{Bruno}, G., {Lewis}, N.~K., {Stevenson}, K.~B., {et~al.} 2018, \aj, 156, 124

\bibitem[{{Cabot} {et~al.}(2019){Cabot}, {Madhusudhan}, {Hawker}, \&
  {Gandhi}}]{2019MNRAS.482.4422C}
{Cabot}, S. H.~C., {Madhusudhan}, N., {Hawker}, G.~A., \& {Gandhi}, S. 2019,
  \mnras, 482, 4422

\bibitem[{{Cabot} {et~al.}(2020){Cabot}, {Madhusudhan}, {Welbanks}, {Piette},
  \& {Gandhi}}]{2020arXiv200107196C}
{Cabot}, S. H.~C., {Madhusudhan}, N., {Welbanks}, L., {Piette}, A., \&
  {Gandhi}, S. 2020, arXiv e-prints, arXiv:2001.07196

\bibitem[{{Casasayas-Barris} {et~al.}(2018){Casasayas-Barris}, {Pall{\'e}},
  {Yan}, {Chen}, {Albrecht}, {Nortmann}, {Van Eylen}, {Snellen}, {Talens},
  {Gonz{\'a}lez Hern{\'a}ndez}, {Rebolo}, \& {Otten}}]{2018A&A...616A.151C}
{Casasayas-Barris}, N., {Pall{\'e}}, E., {Yan}, F., {et~al.} 2018, \aap, 616,
  A151

\bibitem[{{Casasayas-Barris} {et~al.}(2019){Casasayas-Barris}, {Pall{\'e}},
  {Yan}, {Chen}, {Kohl}, {Stangret}, {Parviainen}, {Helling}, {Watanabe},
  {Czesla}, {Fukui}, {Monta{\~n}{\'e}s-Rodr{\'{\i}}guez}, {Nagel}, {Narita},
  {Nortmann}, {Nowak}, {Schmitt}, \& {Zapatero Osorio}}]{2019A&A...628A...9C}
{Casasayas-Barris}, N., {Pall{\'e}}, E., {Yan}, F., {et~al.} 2019, \aap, 628,
  A9

\bibitem[{{Casasayas-Barris} {et~al.}(2020){Casasayas-Barris}, {Palle}, {Yan},
  {Chen}, {Luque}, {Stangret}, {Nagel}, {Zechmeister}, {Oshagh},
  {Sanz-Forcada}, {Nortmann}, {Alonso-Floriano}, {Amado}, {Caballero},
  {Czesla}, {Khalafinejad}, {L\'{o}pez-Puertas}, {L\'{o}pez-Santiago},
  {Molaverdikhani}, {Montes}, {Quirrenbach}, {Reiners}, {Ribas},
  {S\'{a}nchez-L\'{o}pez}, \& {Zapatero Osorio}}]{CasasayasBarris2020}
{Casasayas-Barris}, N., {Palle}, E., {Yan}, F., {et~al.} 2020, \aap, submitted
  (aa37221-19)

\bibitem[{{Cauley} {et~al.}(2018){Cauley}, {Kuckein}, {Redfield}, {Shkolnik},
  {Denker}, {Llama}, \& {Verma}}]{2018AJ....156..189C}
{Cauley}, P.~W., {Kuckein}, C., {Redfield}, S., {et~al.} 2018, \aj, 156, 189

\bibitem[{{Cauley} {et~al.}(2017{\natexlab{a}}){Cauley}, {Redfield}, \&
  {Jensen}}]{2017AJ....153..217C}
{Cauley}, P.~W., {Redfield}, S., \& {Jensen}, A.~G. 2017{\natexlab{a}}, \aj,
  153, 217

\bibitem[{{Cauley} {et~al.}(2017{\natexlab{b}}){Cauley}, {Redfield}, \&
  {Jensen}}]{2017AJ....153..185C}
{Cauley}, P.~W., {Redfield}, S., \& {Jensen}, A.~G. 2017{\natexlab{b}}, \aj,
  153, 185

\bibitem[{{Cauley} {et~al.}(2016){Cauley}, {Redfield}, {Jensen}, \&
  {Barman}}]{2016AJ....152...20C}
{Cauley}, P.~W., {Redfield}, S., {Jensen}, A.~G., \& {Barman}, T. 2016, \aj,
  152, 20

\bibitem[{{Cauley} {et~al.}(2015){Cauley}, {Redfield}, {Jensen}, {Barman},
  {Endl}, \& {Cochran}}]{2015ApJ...810...13C}
{Cauley}, P.~W., {Redfield}, S., {Jensen}, A.~G., {et~al.} 2015, \apj, 810, 13

\bibitem[{{Cauley} {et~al.}(2019){Cauley}, {Shkolnik}, {Ilyin}, {Strassmeier},
  {Redfield}, \& {Jensen}}]{2019AJ....157...69C}
{Cauley}, P.~W., {Shkolnik}, E.~L., {Ilyin}, I., {et~al.} 2019, \aj, 157, 69

\bibitem[{{Cegla} {et~al.}(2016{\natexlab{a}}){Cegla}, {Lovis}, {Bourrier},
  {Beeck}, {Watson}, \& {Pepe}}]{2016A&A...588A.127C}
{Cegla}, H.~M., {Lovis}, C., {Bourrier}, V., {et~al.} 2016{\natexlab{a}}, \aap,
  588, A127

\bibitem[{{Cegla} {et~al.}(2016{\natexlab{b}}){Cegla}, {Oshagh}, {Watson},
  {Figueira}, {Santos}, \& {Shelyag}}]{2016ApJ...819...67C}
{Cegla}, H.~M., {Oshagh}, M., {Watson}, C.~A., {et~al.} 2016{\natexlab{b}},
  \apj, 819, 67

\bibitem[{{Charbonneau} {et~al.}(2002){Charbonneau}, {Brown}, {Noyes}, \&
  {Gilliland}}]{2002ApJ...568..377C}
{Charbonneau}, D., {Brown}, T.~M., {Noyes}, R.~W., \& {Gilliland}, R.~L. 2002,
  \apj, 568, 377

\bibitem[{{Chen} {et~al.}(2017){Chen}, {Pall{\'e}}, {Nortmann}, {Murgas},
  {Parviainen}, \& {Nowak}}]{2017A&A...600L..11C}
{Chen}, G., {Pall{\'e}}, E., {Nortmann}, L., {et~al.} 2017, \aap, 600, L11

\bibitem[{{Christie} {et~al.}(2013){Christie}, {Arras}, \&
  {Li}}]{2013ApJ...772..144C}
{Christie}, D., {Arras}, P., \& {Li}, Z.-Y. 2013, \apj, 772, 144

\bibitem[{{Cincunegui} {et~al.}(2007){Cincunegui}, {D{\'\i}az}, \&
  {Mauas}}]{2007A&A...469..309C}
{Cincunegui}, C., {D{\'\i}az}, R.~F., \& {Mauas}, P.~J.~D. 2007, \aap, 469, 309

\bibitem[{{Cubillos} {et~al.}(2017){Cubillos}, {Fossati}, {Erkaev}, {Malik},
  {Tokano}, {Lendl}, {Johnstone}, {Lammer}, \&
  {Wyttenbach}}]{2017ApJ...849..145C}
{Cubillos}, P.~E., {Fossati}, L., {Erkaev}, N.~V., {et~al.} 2017, \apj, 849,
  145

\bibitem[{{Czesla} {et~al.}(2015){Czesla}, {Klocov{\'a}}, {Khalafinejad},
  {Wolter}, \& {Schmitt}}]{2015A&A...582A..51C}
{Czesla}, S., {Klocov{\'a}}, T., {Khalafinejad}, S., {Wolter}, U., \&
  {Schmitt}, J.~H.~M.~M. 2015, \aap, 582, A51

\bibitem[{{de Mooij} {et~al.}(2017){de Mooij}, {Watson}, \&
  {Kenworthy}}]{2017MNRAS.472.2713D}
{de Mooij}, E. J.~W., {Watson}, C.~A., \& {Kenworthy}, M.~A. 2017, \mnras, 472,
  2713

\bibitem[{{D{\'\i}az} {et~al.}(2007){D{\'\i}az}, {Cincunegui}, \&
  {Mauas}}]{2007MNRAS.378.1007D}
{D{\'\i}az}, R.~F., {Cincunegui}, C., \& {Mauas}, P. J.~D. 2007, \mnras, 378,
  1007

\bibitem[{{Drake} \& {Kashyap}(2010)}]{2010ascl.soft07001D}
{Drake}, J.~J. \& {Kashyap}, V.~L. 2010, {PINTofALE: Package for Interactive
  Analysis of Line Emission}, Astrophysics Source Code Library

\bibitem[{{Eastman} {et~al.}(2013){Eastman}, {Gaudi}, \&
  {Agol}}]{2013PASP..125...83E}
{Eastman}, J., {Gaudi}, B.~S., \& {Agol}, E. 2013, \pasp, 125, 83

\bibitem[{{Ehrenreich}(2010)}]{2010EAS....41..429E}
{Ehrenreich}, D. 2010, in EAS Publications Series, Vol.~41, EAS Publications
  Series, ed. T.~{Montmerle}, D.~{Ehrenreich}, \& A.-M. {Lagrange}, 429--440

\bibitem[{{Fossati} {et~al.}(2018){Fossati}, {Koskinen}, {Lothringer},
  {France}, {Young}, \& {Sreejith}}]{2018ApJ...868L..30F}
{Fossati}, L., {Koskinen}, T., {Lothringer}, J.~D., {et~al.} 2018, \apjl, 868,
  L30

\bibitem[{{Gomes da Silva} {et~al.}(2018){Gomes da Silva}, {Figueira},
  {Santos}, \& {Faria}}]{2018JOSS....3..667G}
{Gomes da Silva}, J., {Figueira}, P., {Santos}, N., \& {Faria}, J. 2018, The
  Journal of Open Source Software, 3, 667

\bibitem[{{Gomes da Silva} {et~al.}(2011){Gomes da Silva}, {Santos}, {Bonfils},
  {Delfosse}, {Forveille}, \& {Udry}}]{2011A&A...534A..30G}
{Gomes da Silva}, J., {Santos}, N.~C., {Bonfils}, X., {et~al.} 2011, \aap, 534,
  A30

\bibitem[{{H{\'e}brard} {et~al.}(2013){H{\'e}brard}, {Collier Cameron},
  {Brown}, {D{\'{\i}}az}, {Faedi}, {Smalley}, {Anderson}, {Armstrong},
  {Barros}, {Bento}, {Bouchy}, {Doyle}, {Enoch}, {G{\'o}mez Maqueo Chew},
  {H{\'e}brard}, {Hellier}, {Lendl}, {Lister}, {Maxted}, {McCormac}, {Moutou},
  {Pollacco}, {Queloz}, {Santerne}, {Skillen}, {Southworth}, {Tregloan-Reed},
  {Triaud}, {Udry}, {Vanhuysse}, {Watson}, {West}, \&
  {Wheatley}}]{2013A&A...549A.134H}
{H{\'e}brard}, G., {Collier Cameron}, A., {Brown}, D.~J.~A., {et~al.} 2013,
  \aap, 549, A134

\bibitem[{{Heng} {et~al.}(2015){Heng}, {Wyttenbach}, {Lavie}, {Sing},
  {Ehrenreich}, \& {Lovis}}]{2015ApJ...803L...9H}
{Heng}, K., {Wyttenbach}, A., {Lavie}, B., {et~al.} 2015, \apjl, 803, L9

\bibitem[{{Hoeijmakers} {et~al.}(2019){Hoeijmakers}, {Ehrenreich}, {Kitzmann},
  {Allart}, {Grimm}, {Seidel}, {Wyttenbach}, {Pino}, {Nielsen}, {Fisher},
  {Rimmer}, {Bourrier}, {Cegla}, {Lavie}, {Lovis}, {Patzer}, {Stock}, {Pepe},
  \& {Heng}}]{2019A&A...627A.165H}
{Hoeijmakers}, H.~J., {Ehrenreich}, D., {Kitzmann}, D., {et~al.} 2019, \aap,
  627, A165

\bibitem[{{Houdebine}(2010)}]{2010MNRAS.403.2157H}
{Houdebine}, E.~R. 2010, \mnras, 403, 2157

\bibitem[{{Huang} {et~al.}(2017){Huang}, {Arras}, {Christie}, \&
  {Li}}]{2017ApJ...851..150H}
{Huang}, C., {Arras}, P., {Christie}, D., \& {Li}, Z.-Y. 2017, \apj, 851, 150

\bibitem[{{Huitson} {et~al.}(2012){Huitson}, {Sing}, {Vidal-Madjar},
  {Ballester}, {Lecavelier des Etangs}, {D{\'e}sert}, \&
  {Pont}}]{2012MNRAS.422.2477H}
{Huitson}, C.~M., {Sing}, D.~K., {Vidal-Madjar}, A., {et~al.} 2012, \mnras,
  422, 2477

\bibitem[{{Hunter}(2007)}]{2007CSE.....9...90H}
{Hunter}, J.~D. 2007, Computing in Science and Engineering, 9, 90

\bibitem[{{Jensen} {et~al.}(2018){Jensen}, {Cauley}, {Redfield}, {Cochran}, \&
  {Endl}}]{2018AJ....156..154J}
{Jensen}, A.~G., {Cauley}, P.~W., {Redfield}, S., {Cochran}, W.~D., \& {Endl},
  M. 2018, \aj, 156, 154

\bibitem[{{Jensen} {et~al.}(2012){Jensen}, {Redfield}, {Endl}, {Cochran},
  {Koesterke}, \& {Barman}}]{2012ApJ...751...86J}
{Jensen}, A.~G., {Redfield}, S., {Endl}, M., {et~al.} 2012, \apj, 751, 86

\bibitem[{{Kausch} {et~al.}(2015){Kausch}, {Noll}, {Smette}, {Kimeswenger},
  {Barden}, {Szyszka}, {Jones}, {Sana}, {Horst}, \&
  {Kerber}}]{2015A&A...576A..78K}
{Kausch}, W., {Noll}, S., {Smette}, A., {et~al.} 2015, \aap, 576, A78

\bibitem[{{Keles} {et~al.}(2019){Keles}, {Mallonn}, {von Essen}, {Carroll},
  {Alexoudi}, {Pino}, {Ilyin}, {Poppenh{\"a}ger}, {Kitzmann}, {Nascimbeni},
  {Turner}, \& {Strassmeier}}]{2019MNRAS.489L..37K}
{Keles}, E., {Mallonn}, M., {von Essen}, C., {et~al.} 2019, \mnras, 489, L37

\bibitem[{{Kirk} {et~al.}(2020){Kirk}, {Alam}, {Lopez-Morales}, \&
  {Zeng}}]{2020arXiv200107667K}
{Kirk}, J., {Alam}, M.~K., {Lopez-Morales}, M., \& {Zeng}, L. 2020, arXiv
  e-prints, arXiv:2001.07667

\bibitem[{{Kirk} {et~al.}(2016){Kirk}, {Wheatley}, {Louden}, {Littlefair},
  {Copperwheat}, {Armstrong}, {Marsh}, \& {Dhillon}}]{2016MNRAS.463.2922K}
{Kirk}, J., {Wheatley}, P.~J., {Louden}, T., {et~al.} 2016, \mnras, 463, 2922

\bibitem[{{Koskinen} {et~al.}(2014){Koskinen}, {Yelle}, {Lavvas}, \&
  {Y-K.~Cho}}]{2014ApJ...796...16K}
{Koskinen}, T.~T., {Yelle}, R.~V., {Lavvas}, P., \& {Y-K.~Cho}, J. 2014, \apj,
  796, 16

\bibitem[{{Lavvas} {et~al.}(2014){Lavvas}, {Koskinen}, \&
  {Yelle}}]{2014ApJ...796...15L}
{Lavvas}, P., {Koskinen}, T., \& {Yelle}, R.~V. 2014, \apj, 796, 15

\bibitem[{{Lendl} {et~al.}(2016){Lendl}, {Delrez}, {Gillon}, {Madhusudhan},
  {Jehin}, {Queloz}, {Anderson}, {Demory}, \& {Hellier}}]{2016A&A...587A..67L}
{Lendl}, M., {Delrez}, L., {Gillon}, M., {et~al.} 2016, \aap, 587, A67

\bibitem[{{Louden} \& {Wheatley}(2015)}]{2015ApJ...814L..24L}
{Louden}, T. \& {Wheatley}, P.~J. 2015, \apjl, 814, L24

\bibitem[{{Louden} {et~al.}(2017){Louden}, {Wheatley}, {Irwin}, {Kirk}, \&
  {Skillen}}]{2017MNRAS.470..742L}
{Louden}, T., {Wheatley}, P.~J., {Irwin}, P. G.~J., {Kirk}, J., \& {Skillen},
  I. 2017, \mnras, 470, 742

\bibitem[{{Madhusudhan}(2019)}]{2019ARA&A..57..617M}
{Madhusudhan}, N. 2019, \araa, 57, 617

\bibitem[{{Mancini} {et~al.}(2017){Mancini}, {Southworth}, {Raia},
  {Tregloan-Reed}, {Molli{\`e}re}, {Bozza}, {Bretton}, {Bruni}, {Ciceri},
  {D'Ago}, {Dominik}, {Hinse}, {Hundertmark}, {J{\o}rgensen}, {Korhonen},
  {Rabus}, {Rahvar}, {Starkey}, {Calchi Novati}, {Figuera Jaimes}, {Henning},
  {Juncher}, {Haugb{\o}lle}, {Kains}, {Popovas}, {Schmidt}, {Skottfelt},
  {Snodgrass}, {Surdej}, \& {Wertz}}]{2017MNRAS.465..843M}
{Mancini}, L., {Southworth}, J., {Raia}, G., {et~al.} 2017, \mnras, 465, 843

\bibitem[{{Mansfield} {et~al.}(2018){Mansfield}, {Bean}, {Oklop{\v{c}}i{\'c}},
  {Kreidberg}, {D{\'e}sert}, {Kempton}, {Line}, {Fortney}, {Henry}, {Mallonn},
  {Stevenson}, {Dragomir}, {Allart}, \& {Bourrier}}]{2018ApJ...868L..34M}
{Mansfield}, M., {Bean}, J.~L., {Oklop{\v{c}}i{\'c}}, A., {et~al.} 2018, \apjl,
  868, L34

\bibitem[{{May} {et~al.}(2018){May}, {Zhao}, {Haidar}, {Rauscher}, \&
  {Monnier}}]{2018AJ....156..122M}
{May}, E.~M., {Zhao}, M., {Haidar}, M., {Rauscher}, E., \& {Monnier}, J.~D.
  2018, \aj, 156, 122

\bibitem[{{McLaughlin}(1924)}]{1924ApJ....60...22M}
{McLaughlin}, D.~B. 1924, \apj, 60, 22

\bibitem[{{Murray-Clay} {et~al.}(2009){Murray-Clay}, {Chiang}, \&
  {Murray}}]{2009ApJ...693...23M}
{Murray-Clay}, R.~A., {Chiang}, E.~I., \& {Murray}, N. 2009, \apj, 693, 23

\bibitem[{{Nortmann} {et~al.}(2018){Nortmann}, {Pall{\'e}}, {Salz},
  {Sanz-Forcada}, {Nagel}, {Alonso-Floriano}, {Czesla}, {Yan}, {Chen},
  {Snellen}, {Zechmeister}, {Schmitt}, {L{\'o}pez-Puertas}, {Casasayas-Barris},
  {Bauer}, {Amado}, {Caballero}, {Dreizler}, {Henning}, {Lamp{\'o}n}, {Montes},
  {Molaverdikhani}, {Quirrenbach}, {Reiners}, {Ribas}, {S{\'a}nchez-L{\'o}pez},
  {Schneider}, \& {Zapatero Osorio}}]{2018Sci...362.1388N}
{Nortmann}, L., {Pall{\'e}}, E., {Salz}, M., {et~al.} 2018, Science, 362, 1388

\bibitem[{{Ochsenbein} {et~al.}(2000){Ochsenbein}, {Bauer}, \&
  {Marcout}}]{2000A&AS..143...23O}
{Ochsenbein}, F., {Bauer}, P., \& {Marcout}, J. 2000, \aaps, 143, 23

\bibitem[{{Ohta} {et~al.}(2005){Ohta}, {Taruya}, \&
  {Suto}}]{2005ApJ...622.1118O}
{Ohta}, Y., {Taruya}, A., \& {Suto}, Y. 2005, \apj, 622, 1118

\bibitem[{{Ohta} {et~al.}(2009){Ohta}, {Taruya}, \&
  {Suto}}]{2009ApJ...690....1O}
{Ohta}, Y., {Taruya}, A., \& {Suto}, Y. 2009, \apj, 690, 1

\bibitem[{{Oklop{\v c}i{\'c}} \& {Hirata}(2018)}]{2018ApJ...855L..11O}
{Oklop{\v c}i{\'c}}, A. \& {Hirata}, C.~M. 2018, \apjl, 855, L11

\bibitem[{{Oshagh} {et~al.}(2013){Oshagh}, {Bou{\'e}}, {Figueira}, {Santos}, \&
  {Haghighipour}}]{2013A&A...558A..65O}
{Oshagh}, M., {Bou{\'e}}, G., {Figueira}, P., {Santos}, N.~C., \&
  {Haghighipour}, N. 2013, \aap, 558, A65

\bibitem[{{Oshagh} {et~al.}(2018){Oshagh}, {Triaud}, {Burdanov}, {Figueira},
  {Reiners}, {Santos}, {Faria}, {Boue}, {D{\'\i}az}, {Dreizler}, {Boldt},
  {Delrez}, {Ducrot}, {Gillon}, {Guzman Mesa}, {Jehin}, {Khalafinejad}, {Kohl},
  {Serrano}, \& {Udry}}]{2018A&A...619A.150O}
{Oshagh}, M., {Triaud}, A.~H.~M.~J., {Burdanov}, A., {et~al.} 2018, \aap, 619,
  A150

\bibitem[{{Parmentier} {et~al.}(2018){Parmentier}, {Line}, {Bean}, {Mansfield},
  {Kreidberg}, {Lupu}, {Visscher}, {D{\'e}sert}, {Fortney}, {Deleuil},
  {Arcangeli}, {Showman}, \& {Marley}}]{2018A&A...617A.110P}
{Parmentier}, V., {Line}, M.~R., {Bean}, J.~L., {et~al.} 2018, \aap, 617, A110

\bibitem[{{Pasquini} \& {Pallavicini}(1991)}]{1991A&A...251..199P}
{Pasquini}, L. \& {Pallavicini}, R. 1991, \aap, 251, 199

\bibitem[{{Pepe} {et~al.}(2010){Pepe}, {Cristiani}, {Rebolo Lopez}, {Santos},
  {Amorim}, {Avila}, {Benz}, {Bonifacio}, {Cabral}, {Carvas}, {Cirami},
  {Coelho}, {Comari}, {Coretti}, {De Caprio}, {Dekker}, {Delabre}, {Di
  Marcantonio}, {D'Odorico}, {Fleury}, {Garc{\'{\i}}a}, {Herreros Linares},
  {Hughes}, {Iwert}, {Lima}, {Lizon}, {Lo Curto}, {Lovis}, {Manescau},
  {Martins}, {M{\'e}gevand}, {Moitinho}, {Molaro}, {Monteiro}, {Monteiro},
  {Pasquini}, {Mordasini}, {Queloz}, {Rasilla}, {Rebord{\~a}o}, {Santana
  Tschudi}, {Santin}, {Sosnowska}, {Span{\`o}}, {Tenegi}, {Udry}, {Vanzella},
  {Viel}, {Zapatero Osorio}, \& {Zerbi}}]{2010SPIE.7735E..0FP}
{Pepe}, F.~A., {Cristiani}, S., {Rebolo Lopez}, R., {et~al.} 2010, in
  \procspie, Vol. 7735, Ground-based and Airborne Instrumentation for Astronomy
  III, 77350F

\bibitem[{{Pino} {et~al.}(2018){Pino}, {Ehrenreich}, {Wyttenbach}, {Bourrier},
  {Nascimbeni}, {Heng}, {Grimm}, {Lovis}, {Malik}, {Pepe}, \&
  {Piotto}}]{2018A&A...612A..53P}
{Pino}, L., {Ehrenreich}, D., {Wyttenbach}, A., {et~al.} 2018, \aap, 612, A53

\bibitem[{{Piskunov} \& {Valenti}(2017)}]{2017A&A...597A..16P}
{Piskunov}, N. \& {Valenti}, J.~A. 2017, \aap, 597, A16

\bibitem[{{Redfield} {et~al.}(2008){Redfield}, {Endl}, {Cochran}, \&
  {Koesterke}}]{2008ApJ...673L..87R}
{Redfield}, S., {Endl}, M., {Cochran}, W.~D., \& {Koesterke}, L. 2008, \apjl,
  673, L87

\bibitem[{{Rossiter}(1924)}]{1924ApJ....60...15R}
{Rossiter}, R.~A. 1924, \apj, 60, 15

\bibitem[{{Ryabchikova} {et~al.}(2015){Ryabchikova}, {Piskunov}, {Kurucz},
  {Stempels}, {Heiter}, {Pakhomov}, \& {Barklem}}]{2015PhyS...90e4005R}
{Ryabchikova}, T., {Piskunov}, N., {Kurucz}, R.~L., {et~al.} 2015, \physscr,
  90, 054005

\bibitem[{{Salz} {et~al.}(2018){Salz}, {Czesla}, {Schneider}, {Nagel},
  {Schmitt}, {Nortmann}, {Alonso-Floriano}, {L{\'o}pez-Puertas}, {Lamp{\'o}n},
  {Bauer}, {Snellen}, {Pall{\'e}}, {Caballero}, {Yan}, {Chen}, {Sanz-Forcada},
  {Amado}, {Quirrenbach}, {Ribas}, {Reiners}, {B{\'e}jar}, {Casasayas-Barris},
  {Cort{\'e}s-Contreras}, {Dreizler}, {Guenther}, {Henning}, {Jeffers},
  {Kaminski}, {K{\"u}rster}, {Lafarga}, {Lara}, {Molaverdikhani}, {Montes},
  {Morales}, {S{\'a}nchez-L{\'o}pez}, {Seifert}, {Zapatero Osorio}, \&
  {Zechmeister}}]{2018A&A...620A..97S}
{Salz}, M., {Czesla}, S., {Schneider}, P.~C., {et~al.} 2018, \aap, 620, A97

\bibitem[{{Seidel} {et~al.}(2020){Seidel}, {Ehrenreich}, {Pino}, {Bourrier},
  {Lavie}, {Allart}, {Wyttenbach}, \& {Lovis}}]{2020A&A...633A..86S}
{Seidel}, J.~V., {Ehrenreich}, D., {Pino}, L., {et~al.} 2020, \aap, 633, A86

\bibitem[{{Seidel} {et~al.}(2019){Seidel}, {Ehrenreich}, {Wyttenbach},
  {Allart}, {Lendl}, {Pino}, {Bourrier}, {Cegla}, {Lovis}, {Barrado},
  {Bayliss}, {Astudillo-Defru}, {Deline}, {Fisher}, {Heng}, {Joseph}, {Lavie},
  {Melo}, {Pepe}, {S{\'e}gransan}, \& {Udry}}]{2019A&A...623A.166S}
{Seidel}, J.~V., {Ehrenreich}, D., {Wyttenbach}, A., {et~al.} 2019, \aap, 623,
  A166

\bibitem[{{Shporer} \& {Brown}(2011)}]{2011ApJ...733...30S}
{Shporer}, A. \& {Brown}, T. 2011, \apj, 733, 30

\bibitem[{{Sing} {et~al.}(2008){Sing}, {Vidal-Madjar}, {D{\'e}sert},
  {Lecavelier des Etangs}, \& {Ballester}}]{2008ApJ...686..658S}
{Sing}, D.~K., {Vidal-Madjar}, A., {D{\'e}sert}, J.-M., {Lecavelier des
  Etangs}, A., \& {Ballester}, G. 2008, \apj, 686, 658

\bibitem[{{Smette} {et~al.}(2015){Smette}, {Sana}, {Noll}, {Horst}, {Kausch},
  {Kimeswenger}, {Barden}, {Szyszka}, {Jones}, {Gallenne}, {Vinther},
  {Ballester}, \& {Taylor}}]{2015A&A...576A..77S}
{Smette}, A., {Sana}, H., {Noll}, S., {et~al.} 2015, \aap, 576, A77

\bibitem[{{Snellen} {et~al.}(2008){Snellen}, {Albrecht}, {de Mooij}, \& {Le
  Poole}}]{2008A&A...487..357S}
{Snellen}, I.~A.~G., {Albrecht}, S., {de Mooij}, E.~J.~W., \& {Le Poole}, R.~S.
  2008, \aap, 487, 357

\bibitem[{{Snellen} {et~al.}(2010){Snellen}, {de Kok}, {de Mooij}, \&
  {Albrecht}}]{2010Natur.465.1049S}
{Snellen}, I.~A.~G., {de Kok}, R.~J., {de Mooij}, E.~J.~W., \& {Albrecht}, S.
  2010, \nat, 465, 1049

\bibitem[{{Southworth}(2008)}]{2008MNRAS.386.1644S}
{Southworth}, J. 2008, \mnras, 386, 1644

\bibitem[{{Southworth}(2011)}]{2011MNRAS.417.2166S}
{Southworth}, J. 2011, \mnras, 417, 2166

\bibitem[{{Spake} {et~al.}(2018){Spake}, {Sing}, {Evans}, {Oklop{\v c}i{\'c}},
  {Bourrier}, {Kreidberg}, {Rackham}, {Irwin}, {Ehrenreich}, {Wyttenbach},
  {Wakeford}, {Zhou}, {Chubb}, {Nikolov}, {Goyal}, {Henry}, {Williamson},
  {Blumenthal}, {Anderson}, {Hellier}, {Charbonneau}, {Udry}, \&
  {Madhusudhan}}]{2018Natur.557...68S}
{Spake}, J.~J., {Sing}, D.~K., {Evans}, T.~M., {et~al.} 2018, \nat, 557, 68

\bibitem[{{Turner} {et~al.}(2020){Turner}, {de Mooij}, {Jayawardhana}, {Young},
  {Fossati}, {Koskinen}, {Lothringer}, {Karjalainen}, \&
  {Karjalainen}}]{2020ApJ...888L..13T}
{Turner}, J.~D., {de Mooij}, E. J.~W., {Jayawardhana}, R., {et~al.} 2020,
  \apjl, 888, L13

\bibitem[{{Valenti} \& {Piskunov}(1996)}]{1996A&AS..118..595V}
{Valenti}, J.~A. \& {Piskunov}, N. 1996, \aaps, 118, 595

\bibitem[{{{\v{Z}}{\'a}k} {et~al.}(2019){{\v{Z}}{\'a}k}, {Kab{\'a}th},
  {Boffin}, {Ivanov}, \& {Skarka}}]{2019AJ....158..120Z}
{{\v{Z}}{\'a}k}, J., {Kab{\'a}th}, P., {Boffin}, H. M.~J., {Ivanov}, V.~D., \&
  {Skarka}, M. 2019, \aj, 158, 120

\bibitem[{{Wyttenbach} {et~al.}(2015){Wyttenbach}, {Ehrenreich}, {Lovis},
  {Udry}, \& {Pepe}}]{2015A&A...577A..62W}
{Wyttenbach}, A., {Ehrenreich}, D., {Lovis}, C., {Udry}, S., \& {Pepe}, F.
  2015, \aap, 577, A62

\bibitem[{{Wyttenbach} {et~al.}(2017){Wyttenbach}, {Lovis}, {Ehrenreich},
  {Bourrier}, {Pino}, {Allart}, {Astudillo-Defru}, {Cegla}, {Heng}, {Lavie},
  {Melo}, {Murgas}, {Santerne}, {S{\'e}gransan}, {Udry}, \&
  {Pepe}}]{2017A&A...602A..36W}
{Wyttenbach}, A., {Lovis}, C., {Ehrenreich}, D., {et~al.} 2017, \aap, 602, A36

\bibitem[{{Yan} {et~al.}(2015){Yan}, {Fosbury}, {Petr-Gotzens}, {Zhao}, \&
  {Pall{\'e}}}]{2015A&A...574A..94Y}
{Yan}, F., {Fosbury}, R.~A.~E., {Petr-Gotzens}, M.~G., {Zhao}, G., \&
  {Pall{\'e}}, E. 2015, \aap, 574, A94

\bibitem[{{Yan} \& {Henning}(2018)}]{2018NatAs...2..714Y}
{Yan}, F. \& {Henning}, T. 2018, Nature Astronomy, 2, 714

\bibitem[{{Yan} {et~al.}(2017){Yan}, {Pall{\'e}}, {Fosbury}, {Petr-Gotzens}, \&
  {Henning}}]{2017A&A...603A..73Y}
{Yan}, F., {Pall{\'e}}, E., {Fosbury}, R.~A.~E., {Petr-Gotzens}, M.~G., \&
  {Henning}, T. 2017, \aap, 603, A73

\bibitem[{{Yana Galarza} {et~al.}(2019){Yana Galarza}, {Mel{\'e}ndez},
  {Lorenzo-Oliveira}, {Valio}, {Reggiani}, {Carlos}, {Ponte}, {Spina},
  {Haywood}, \& {Gandolfi}}]{2019MNRAS.490L..86Y}
{Yana Galarza}, J., {Mel{\'e}ndez}, J., {Lorenzo-Oliveira}, D., {et~al.} 2019,
  \mnras, 490, L86

\bibitem[{{Yelle}(2004)}]{2004Icar..170..167Y}
{Yelle}, R.~V. 2004, \icarus, 170, 167

\end{thebibliography}


\begin{appendix}

\section{Joint fit of the Rossiter-McLaughlin effect in the radial velocity time series}
\label{sec:rmfit}

To determine whether any offset exists in the calculated mid-transit times based on the literature ephemeris, we jointly fit the stellar RV time series of the three nights for the Rossiter-McLaughlin effect. We used the EXOFAST \citep{2013PASP..125...83E} parameterization of the \citet{2005ApJ...622.1118O} formulae and adopted a circular orbit. The free parameters are: offset to expected mid-transit time $\Delta T_C$, systemic velocity $\gamma$, linear limb-darkening coefficient $u_1$, stellar radial velocity semi-amplitude $K_\star$, projected stellar surface velocity $v\sin i_\star$, and projected spin-orbit angle $\lambda$. The other parameters are fixed to literature values listed in Table~\ref{tab:param}, including orbital period $P$, orbital inclination $i$, scaled semi-major axis $a/R_\star$, and planet-to-star radius ratio $R_\mathrm{p}/R_\star$. The three nights were forced to share the same $\Delta T_C$, which assumed that the change of orbital period is negligible during the two weeks that cover our three transit observations. 

\begin{figure}
\centering
\includegraphics[width=\linewidth]{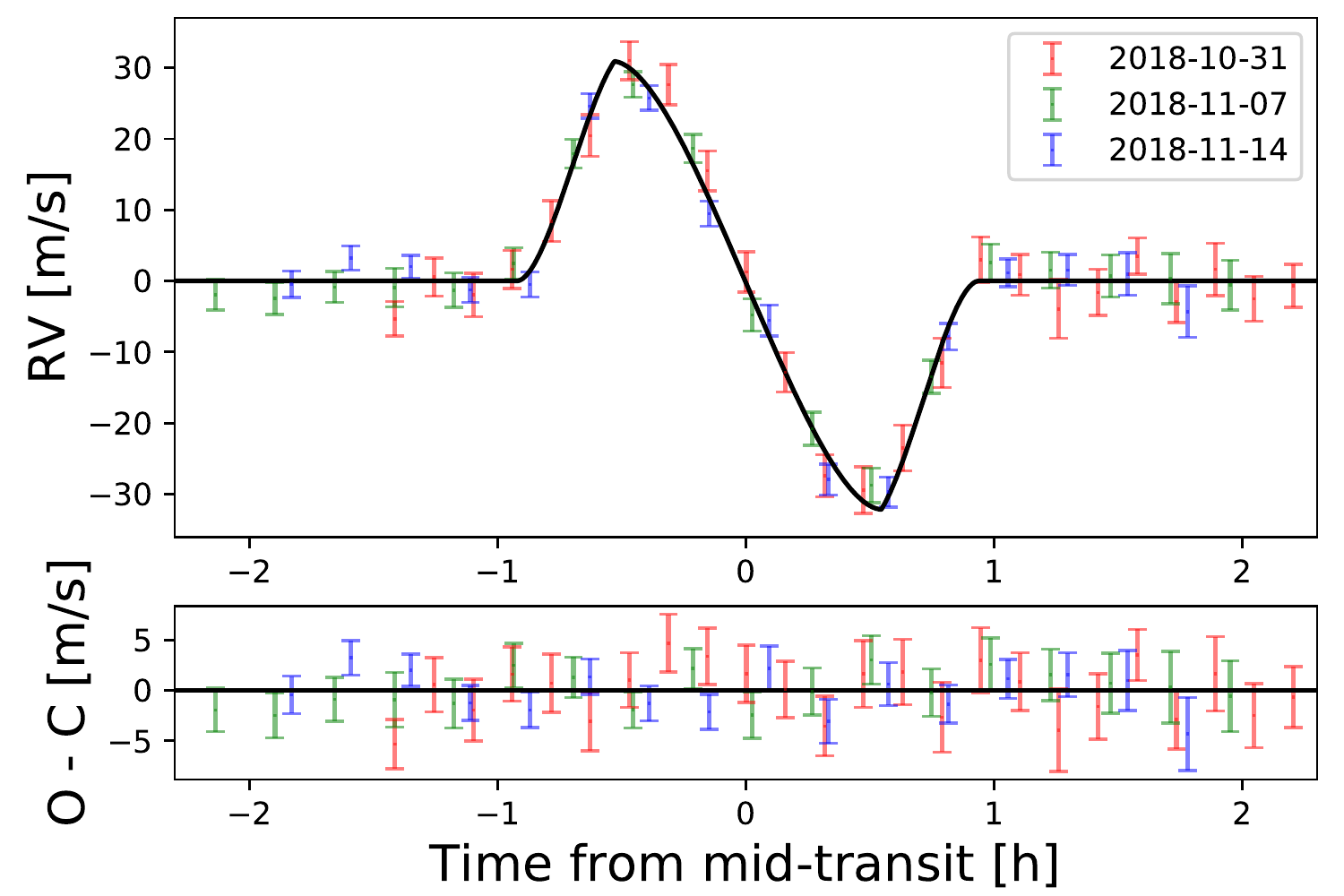}
\caption{Top panel: measured Rossiter-McLaughlin effect of WASP-52b in the stellar RV time series. It consists of three nights in different colors. The solid line is the best-fit Rossiter-McLaughlin model. Bottom panel: the best-fit residuals.}
\label{fig:rmfit}
\end{figure}

The fitting results do not reveal any significant offset ($\Delta T_C=0.00031\pm 0.00032$~days). As shown in Figure~\ref{fig:rmfit}, the Rossiter-McLaughlin effect is unambiguously detected at high significance. We obtained systemic velocities of $-821.16\pm 0.66$~m\,s$^{-1}$, $-843.39\pm 0.63$~m\,s$^{-1}$,  $-835.69\pm 0.56$~m\,s$^{-1}$ for the transits on the nights of 31 Oct, 7 and 14 Nov, respectively. We derived a stellar RV semi-amplitude of 82.8 $\pm$ 2.1~m\,s$^{-1}$ solely based on the Rossiter-McLaughlin analysis, which agrees well with the value derived from full-phase RV analysis \citep[$K_\star=84.3\pm 3.0$~m\,s$^{-1}$;][]{2013A&A...549A.134H}. We also measured a projected rotation velocity of $2.62\pm 0.074$~km\,s$^{-1}$ for the star, which is marginally below the value from our spectral analysis on the coaaded ESPRESSO spectrum: $3.13\pm 0.42$~km\,s$^{-1}$. The discrepancy in the $v\sin i$ values derived from the Rossiter-McLaughlin effect or from the spectral line broadening has been noticed in some studies \citep[e.g.,][]{2017MNRAS.464..810B,2018A&A...619A.150O}, which might be caused by underestimated uncertainties from any of the two methods \citep{2017MNRAS.464..810B}, stellar differential rotation \citep{2012ApJ...757...18A,2016A&A...588A.127C}, stellar convective blueshift and granulation \citep{2011ApJ...733...30S,2016ApJ...819...67C}, gravitational microlensing \citep{2013A&A...558A..65O}, or exo-rings \citep{2009ApJ...690....1O,2017MNRAS.472.2713D,2018A&A...609A..21A}. Our measured projected spin-orbit angle is $1.1^{\circ}\pm 1.1^{\circ}$, which indicates that the projected planetary orbit normal is well-aligned with the stellar spin axis. 

A detailed ``reloaded'' Rossiter-McLaughlin analysis will be presented in another paper (Cegla et al. in preparation), which is out of the scope of this paper. The reloaded RM technique \citep{2016A&A...588A.127C} uses the planet as a probe to isolate the local CCFs from the regions successively occulted during the transit, with no particular assumptions about the shape of the intrinsic stellar photospheric lines. It scans the RV of the stellar surface along the transit chord and allows true 3D obliquity to be inferred. This technique has been implemented on several transiting planetary systems, including HD 189733 \citep{2016A&A...588A.127C}, WASP-8 \citep{2017A&A...599A..33B}, WASP-49 \citep{2017A&A...602A..36W}, GJ 436 \citep{2018Natur.553..477B} and WASP-121 \citep{2020arXiv200106836B}.

\end{appendix}

\end{document}